\documentclass[prb,preprint,showpacs]{revtex4}

\usepackage{graphicx} 
\usepackage{dcolumn} 
\usepackage{amsmath}
\newcommand{\func}[3]{\ensuremath{\operatorname{#1}_{#2}\left(#3\right)}}

\begin{document}

\preprint{APS/123-QED}

\title{Modeling molecular and ionic absolute solvation free 
energies with quasi-chemical theory bounds}


\author{David M. Rogers}
\affiliation{Department of Chemistry, University of Cincinnati, \\
Cincinnati, OH 45221-0172}
\author{Thomas L. Beck}
\email{thomas.beck@uc.edu}
\affiliation{Departments of Chemistry and Physics, University of Cincinnati, \\
Cincinnati, OH 45221-0172}

\date{\today}

\begin{abstract}
A recently developed statistical mechanical Quasi-Chemical 
Theory (QCT) has led to significant insights into solvation phenomena for both 
hydrophilic and hydrophobic solutes.  The QCT exactly partitions solvation free energies 
into three components: 1) inner-shell 
chemical, 
2) outer-shell packing,  and 3) outer-shell long-ranged contributions.  In this paper, we discuss 
efficient methods for computing each of the three
parts of the free energy.  A Bayesian estimation approach is developed to 
compute the inner-shell chemical and outer-shell packing contributions.  
We derive upper and lower bounds on the outer-shell long-ranged portion of the free energy by expressing this component in two equivalent ways.  Local, high energy contacts between solute and solvent are eliminated by spatial conditioning in this free energy piece, leading to near-Gaussian distributions of solute-solvent interactions energies.  Thus, the 
average of the two mean-field bounds yields an accurate and efficient 
free energy estimate.  
Aqueous solvation free energy results are presented for several solutes, including
methane, perfluoromethane, water, and the sodium and chloride ions.  The results 
demonstrate the accuracy and efficiency of the methods.  The 
approach should prove useful 
in computing solvation free energies 
in inhomogeneous, restricted environments. 
\end{abstract}
\pacs{82.60.Lf,87.16.A-,61.20.Ja,64.70.qd,64.75.Bc}
\keywords{Quasi-chemical Theory,Free Energy,Solvation,Bayesian}

\maketitle

\section{Introduction}{\label{sec_Intro}}

Free energy calculations are a central challenge in modern 
statistical mechanics.\cite{febook}  Accurate 
evaluations of free energies are crucial for understanding
a wide range of chemical and physical phenomena, 
including chemical reactions in solution,\cite{wjorg89,pkoll93,pkoll96,jflor98,hhu08} solvation free 
energies,\cite{wjorg89,pkoll93,pkoll96,mshirts03,ydeng04,mshirts05,fytre06,glam06,dmob07} potentials of mean force between
complex molecular species in water,\cite{wjorg90,dast08} acid-base equilibria,\cite{tsimonson07,cchip07b} ligand/drug binding interactions,\cite{wjorg89,pkoll96,wjorg04,jwang06,dmob07a,ydeng08} and 
phase equilibria.\cite{mshel07a}  Free energies are composed of enthalpic and entropic pieces, 
both of which may be more difficult to estimate than the free energy itself
during computer simulations.\cite{pkoll93,nlu03}  Due to the high sensitivity of computed free energies,
enthalpies, and entropies to 
force field variations, and the possibility of direct and accurate
experimental testing of the computed results, 
calculations of these thermodynamic quantities provide a proving ground for the development of viable
molecular models.\cite{ARCC3fem,case03,wjorg05,glam06}

A wide range of theoretical and numerical methods has been developed 
for assessing 
free energy changes.\cite{fytre06a}  The standard toolbox of free energy calculations
includes the thermodynamic integration (TI) and free energy perturbation (FEP)
techniques.  The theory behind these methods was developed long ago by
Kirkwood,\cite{jkirk35} Landau and Lifshitz,\cite{LL} Zwanzig,\cite{rwzwan54} and others.  More recent developments
have included the histogram overlap,\cite{jvall72} Bennett acceptance ratio (BAR),\cite{chbenn76} importance (or umbrella) sampling,\cite{gmtorr77} 
adaptive biasing force (ABF),\cite{edarv07}
weighted histogram (WHAM),\cite{skum92} replica exchange,\cite{wnad07} resolution exchange,\cite{elym06} metadynamics,\cite{laio02} and non-equilibrium work 
methods.\cite{cjarz97,ghum07} Ref.~\onlinecite{febook} offers a thorough overview of the fundamentals of 
these and other modern approaches to
free energy calculations.  

Often a coupling parameter is introduced that, when 
changed from zero to one, gradually links the initial and final states 
along some path.  This partitioning of the problem 
along a coupling path is typically referred to as stratification or staging.\cite{cchip07}
The FEP approach samples the exponential of the energy difference between two steps in an alchemical transformation,\cite{cchip07} while TI samples the analytic derivative of the Hamiltonian with respect 
to the coupling parameter for a given value of that parameter.\cite{edarv07}  
A key feature of these traditional free energy methods is that
there must be robust sampling of the phase spaces along 
the coupling parameter transition; the smaller are the coupling parameter increments, 
the larger are the overlaps of the phase spaces sampled throughout the transition.  Extensive
work has gone into optimizing algorithms to enhance these phase space 
overlaps.\cite{berne97,nlu99,nlu03a,mshell07,dmin08}  
In the present paper, we 
develop an alternative approach based on a {\em spatial} stratification of the free energy contributions,
and apply the methods to the computation of absolute hydration free energies.  

As suggested above, FEP calculations invariably encounter the same fundamental difficulty, namely that of sampling the distribution of the energy change.\cite{mshirts05,cchip07}  For the case of solvation free energy considered here, and without any stratification, this energy change is the interaction energy between the solvent and its environment.   Taking the exponent of this quantity, however, makes the estimate highly sensitive to the contributions from tail regions of the interaction energy probability distribution.  The probability of sampling these tail regions during standard simulations may be low, leading to poor statistics, and results from naive applications of the exponential averaging formula may thus incur substantial errors.  In addition, the averaging contains a 
systematic bias.\cite{dzuck02,ourbook,nlu07}

The above point can be illustrated with the histogram overlap method, Fig.~\ref{fig:attpi}.\cite{ourbook,lpratt07}  That method takes advantage of data from two simulations (initial and final states) by noting that the logarithm of the ratio of these energy distributions is related to the free energy difference.  We display the interaction energy distributions for the coupled and uncoupled limits for methane solvation in water.  In the figure, it can be seen that the probability distribution of the solute-solvent interaction energy undergoes significant changes during the solvation process.  Those configurations whose energies are most important for characterizing the solvated state occupy only a small fraction of the total configurational space of the uncoupled system.  For the displayed case, it is clear that, even though methane is a reasonably small-sized molecule, large simulation times are required to adequately sample the low-energy tail of the uncoupled distribution (as it occupies only 0.2\% of the sample space).  An exponential averaging formula does not work for the fully coupled case, since the high-energy, exponential tail is not adequately sampled during simulations; direct implementation of the exponential averaging formula leads to a 16 kJ/mol error in the free energy. Larger cases such as CF$_4$ become even more problematic,\cite{dasth07} and no detectable overlap would occur for a small peptide, for example.

Recognizing this fundamental difficulty, some free energy methods attempt to circumvent the overlap problem by a reformulation designed to minimize the error.  Stratification,\cite{cchip07} importance sampling,\cite{mshell07} and the multicanonical,\cite{bber92} BAR,\cite{chbenn76} and WHAM\cite{skum92} methods are efforts in that direction.  Utilization of Tsallis statistics\cite{berne97,iandri07} is another possibility in which the canonical probability distribution is re-expressed in a form that has the Boltzmann distribution as a particular limit, effectively smoothing out the potential surface. This allows the higher-energy regions to be reached more frequently and can improve the sampling of the tail regions at the expense of less detailed exploration of local minima. 

Other approaches include cumulant expansions of the free energy\cite{ghumm962} and high-order TI formulas.\cite{ghumm961}  A cumulant expansion is a special case of fitting the energy probability distribution to an approximate analytic form.  Cumulant expansions to second order may be particularly effective in some circumstances, since the two-moment formula is exact in the case that the energy is Gaussian-distributed.  Combinations of accurate, high-order TI integration formulas with cumulant expansions (to obtain derivatives in a Taylor expansion of the free energy) have led to very efficient free energy estimates.\cite{ghumm961}

The Potential Distribution Theorem (PDT) provides another view of solvation thermodynamics that can be helpfully employed in free energy calculations.\cite{bwido63,ourbook,lpratt07}  It results in a near-local partition function expression for the excess chemical potential. That excess chemical potential is the solvation free energy sought here.  Following the original formulation by 
Widom,\cite{bwido63} the PDT has found wide application in models of small molecule solvation via the test particle method.\cite{dfren96}  More recently, the PDT has been further expanded as a general statistical mechanical approach for molecular liquids.\cite{ourbook}  Part of that conceptual expansion has involved the development of a quasi-chemical  theory (QCT)\cite{lrprat98,lrpwater03,dasth03,lpratt07,dasth07,ourbook} that exactly partitions the free energy into inner-shell (IS) and outer-shell (OS) components.  The exact expression can then be manipulated to make physical approximations for the IS and OS parts.  The OS portion of the free energy can be further decomposed into packing and long-ranged contributions.\cite{dasth07,dast08} The term `long-ranged' used in this context indicates the free energy for interaction of the solute with the solvent, with solvent molecules excluded from the inner-shell region.

The spatial partitioning may be utilized to perform, for example, {\em ab initio} calculations for the IS domain, while making classical mechanical or even continuum approximations for the OS region.\cite{dasth03,ourbook}   The QCT has been found especially effective for understanding ion solvation phenomena,\cite{dasth03,dast03,dast04,sremp04,ourbook} 
leading to highly accurate estimates of the free energies.
Recently, quasi-chemical theory has been further applied to the calculation of the aqueous solvation free energies of water\cite{apali06,jshah07} and small nonpolar molecules such as 
perfluoromethane (CF$_4$)\cite{dasth07} and methane.\cite{dast08}
These studies have found that the absence of close-contact repulsion in the OS long-ranged part of the free energy induced by the QCT partitioning leads to near-Gaussian energy distributions.  In this paper we exploit that observation to develop efficient simulation methods for computing this portion of the free energy. A novel statistical procedure for estimation of the IS and OS packing contributions is also presented, leading to a robust procedure for calculating absolute solvation free energies directly from molecular dynamics sampling.  

The structure of the paper is as follows.  Section~\ref{sec_QCT} introduces the theoretical motivation behind the QCT and the sampling strategies employed, and derives bounds on the OS long-ranged part of the free energy.  Section~\ref{sec_Bayes} develops a Bayesian approach for computing the IS free energy and the OS packing contribution. Section~\ref{sec_MD} illustrates how a continuous model repulsive potential can be included to allow for molecular dynamics sampling in the free energy computations, and Section~\ref{sec_Simulations} discusses details of the simulations reported here.  Section~\ref{sec_Results}
presents results of absolute solvation free energy calculations for methane, perfluoromethane, TIP3P water, and the sodium and chloride ions.  The final section presents our conclusions and discusses potential applications to more advanced free energy problems.

\section{Quasi-chemical theory}
\label{sec_QCT}

In this section, we review the Potential Distribution Theorem (PDT) and its quasi-chemical theory (QCT) form to set the stage for developing efficient computational approaches for calculating the free energy via spatial partitioning; see Refs.~\onlinecite{ourbook,lpratt07} for further discussion.  
The PDT expression for the chemical potential 
of chemical species is
\begin{equation}
\beta\mu_{\alpha} = \ln\left[\rho_{\alpha}({\bf r})\Lambda^3_{\alpha}
/q_{\alpha}^{\mathrm{int}}\right] - 
  \ln \langle\langle \exp(-\beta\Delta U_{\alpha}) | {\bf r} \rangle\rangle_0 .
\end{equation}
We typically express the chemical potentials in unitless form, $\beta\mu_{\alpha}$, where $\beta = 1/{k_B T}$ ($k_B$ is the Boltzmann constant and $T$ is the temperature). The first term on the rhs is the ideal chemical potential involving the
molecular center-of-mass number density $\rho_{\alpha}({\bf r})$, the thermal deBroglie wavelength 
$\Lambda_{\alpha}$, and the 
gas phase partition function $q_{\alpha}^{\mathrm{int}}$ for the internal states of the molecule (vibrations,
rotations, and electronic).  The second term is the excess chemical potential
evaluated at the center-of-mass point ${\bf r}$.  The double brackets with zero 
subscript imply that the $N$ solvent molecules and the solute are sampled independently;
then the two independent subsystems are superimposed and the Boltzmann
factor of the interaction energy ($\Delta U_{\alpha} = U_{N+1} - U_N$) is averaged.
It is clear that, if the solute molecule is large, in practice most superimposed subsystems
will lead to significant overlap of the hard molecular cores, even though the 
formula is rigorously correct.  This is one motivation for the development of the 
partitioning scheme of Quasi-Chemical Theory (QCT.)\cite{lrprat98,ourbook,lpratt07}
 
In this paper, the chemical potential sought is for a single solute in dilute solution, 
so no $\alpha$ label
is necessary.  
We consider solute molecules either of the united atom type or with 
fixed internal coordinates, dissolved in a homogeneous liquid (water).  Thus, we also do not need 
to consider the sampling of the solute internal coordinates, and we can drop the
position $\bf{r}$. The PDT then reduces to
\begin{equation}
\beta\mu = \ln\left[\rho\Lambda^3\right] - 
  \ln \langle \exp(-\beta\Delta U) \rangle_0 .
\end{equation}

Our focus will be on the excess chemical potential, that is the contribution over and above 
the ideal term:
\begin{equation}
\beta\mu^{\mathrm{ex}} = - \ln \langle \exp(-\beta\Delta U)  \rangle_0  .
\end{equation}
The average on the right hand side can be expressed as 
\begin{equation}
\langle \mathrm{e}^{-\beta\Delta U} \rangle_0 = \frac
{\int dx^N \mathrm{e}^{-\beta U_N}\mathrm{e}^{-\beta\Delta U }}
{\int dx^N \mathrm{e}^{-\beta U_N}} ,
\label{insertprob}
\end{equation}
where $U_N$ represents the interaction energy for the $N$ solvent molecules, and 
$\Delta U$ is the interaction energy of the solute with the solvent.  The above notation
is a schematic in which the integration over all $3N$ solvent molecular degrees of freedom is
represented by $\int dx^N$ while the solute center-of-mass is held fixed.

Alternatively, we can express Eq.~(\ref{insertprob}) as  
\begin{equation}
\langle \mathrm{e}^{-\beta\Delta U} \rangle_0 = 
\mathrm{e}^{-\beta\mu^{ex} } =
\int d\epsilon P^{(0)} (\epsilon) \mathrm{e}^{-\beta\epsilon} ,
\label{DOS}
\end{equation}
where
\begin{equation}
P^{(0)} (\epsilon) = \langle \delta (\epsilon - \Delta U) \rangle_0 
\label{DOS1}
\end{equation}
is the distribution of the interaction energies in the uncoupled case.

Similarly, we can say  
\begin{equation}
\langle \mathrm{e}^{\beta\Delta U} \rangle = 
\mathrm{e}^{\beta\mu^{\mathrm{ex}} } =
\int d\epsilon P (\epsilon) \mathrm{e}^{\beta\epsilon} ,
\label{DOS2}
\end{equation}
where
\begin{equation}
P (\epsilon) = \langle \delta (\epsilon - \Delta U) \rangle  . 
\label{DOS3}
\end{equation}
The lack of a zero subscript for the average in Eq.~(\ref{DOS3}) implies that the solute
is included in the sampling.
This is the inverse form of the PDT.  There is a simple relationship between the 
coupled and uncoupled binding energy distributions:\cite{ourbook}
\begin{equation}
P (\epsilon) = \mathrm{e}^{-\beta(\epsilon-\mu^{\mathrm{ex}})} P^{(0)}(\epsilon) .
\label{DOS4}
\end{equation}
Thus, if one of the two distributions were to assume a Gaussian form, then so would the other. At another extreme, if one distribution were roughly constant, then the other would be near-exponential.

We can partition the excess chemical potential by inserting unity 
into Eq.~(\ref{insertprob}) in the form
\begin{equation}
1= \frac
{\int dx^N \mathrm{e}^{-\beta U_N}\mathrm{e}^{-\beta\Delta U_{\mathrm{HS}} (\lambda) } }
{\int dx^N \mathrm{e}^{-\beta U_N}\mathrm{e}^{-\beta\Delta U_{\mathrm{HS}} (\lambda) } }
\frac
{\int dx^N \mathrm{e}^{-\beta U_N}\mathrm{e}^{-\beta\Delta U_{\mathrm{HS}} (\lambda) } \mathrm{e}^{-\beta\Delta U}}
{\int dx^N \mathrm{e}^{-\beta U_N}\mathrm{e}^{-\beta\Delta U_{\mathrm{HS}} (\lambda)} \mathrm{e}^{-\beta\Delta U}}  ,
\end{equation}
where $\Delta U_{\mathrm{HS}} (\lambda)$ is a hard-sphere potential and $\lambda$ is the distance from the center of the hard sphere to the center of the nearest solvent molecule.  The radius 
$\lambda$ divides the system into inner-shell (IS) and outer-shell (OS) domains.   
It is typically chosen so that the solute/solvent radial distribution function at that distance
has appreciable nonzero values, so as to include some solvent 
molecules during thermal sampling; distances out to the first minimum in the 
radial distribution function might be considered.  Thus the IS domain includes the solute and some
solvent molecules directly interacting with the solute, and the OS region includes all of the rest 
of the solvent.   

After a slight rearrangement 
\begin{equation}
\langle \mathrm{e}^{-\beta\Delta U} \rangle_0 = 
\frac{p_0 (\lambda)}{x_0 (\lambda)} \langle \mathrm{e}^{-\beta\Delta U} \rangle_{\lambda} .
\label{eq:qct}
\end{equation}
Here
\begin{equation}
p_0 (\lambda) = \langle\mathrm{e}^{-\beta\Delta U_{\mathrm{HS}} (\lambda) } \rangle_0 ,
\label{p_0}
\end{equation}
and 
\begin{equation}
x_0 (\lambda) = \langle\mathrm{e}^{-\beta\Delta U_{\mathrm{HS}} (\lambda)} \rangle .
\label{x_0}
\end{equation}
The sampling leading to the average labeled by $\lambda$ in Eq.~(\ref{eq:qct}) includes the hard particle with size specified by $\lambda$.  Also, the solute is included in 
 the sampling that yields $x_0 (\lambda)$:
\begin{equation}
x_0 (\lambda) = \langle\mathrm{e}^{-\beta\Delta U_{\mathrm{HS}} (\lambda) } \rangle
= \frac
{\int dx^N \mathrm{e}^{-\beta U_N}\mathrm{e}^{-\beta\Delta U  } \mathrm{e}^{-\beta\Delta U_{\mathrm{HS}} (\lambda) } }
{\int dx^N \mathrm{e}^{-\beta U_N}\mathrm{e}^{-\beta\Delta U  } }  .
\label{eq:x_0form}
\end{equation}
The quantity that is averaged in Eq.~(\ref{eq:x_0form}),
 $\exp [-\beta\Delta U_{\mathrm{HS}} (\lambda)]$, is an `indicator function' that
is one if there are no overlaps with solvent molecules and zero if overlaps exist.\cite{ourbook} 
The term $x_0 (\lambda)$ is the probability of observing no solvent centers within the IS region
while the solute molecule is situated at the center of the of the inner shell.  The term
$p_0 (\lambda)$ is the probability that no solvent centers are located within the IS domain with
no solute in that domain.  

Thus the excess chemical potential can be written as
\begin{equation}
\beta\mu^{\mathrm{ex}} = \ln x_0 (\lambda) - \ln p_0 (\lambda) - \ln \langle \mathrm{e}^{-\beta\Delta U}
 \rangle_{\lambda} .
\label{mu_QCT}
\end{equation}
It is important to note that this spatial partitioning of the free energy is exact; the introduction of 
a hard-sphere particle is an artifice to enact the partitioning.  The result for the excess
chemical potential is independent of the choice of $\lambda$, as we will see in our
modeling results, and has been observed in previous calculations.\cite{apali06}   
There may be judicious choices for $\lambda$, though, that can lead
to helpful approximations that enhance the efficiency and utility of the theory.  

The first term on the rhs of Eq.~(\ref{mu_QCT}) is the IS contribution to the excess chemical 
potential, $\beta\mu_{\mathrm{IS}}^{\mathrm{ex}}(\lambda)$.  
It corresponds to minus the work necessary to 
move the nearby solvent molecules from
direct contact with the solute to outside the IS domain.  Thus, this term can also be viewed as
a chemical contribution that is, roughly speaking, the free energy to bind the nearest solvent 
molecules to the solute.    The second term is the work to grow in a cavity 
of size $\lambda$ in the solvent.
This is the OS packing contribution  $\beta\mu_{\mathrm{OS,HS}}^{\mathrm{ex}}(\lambda)$.  The final term is the free energy of interaction of the solute
with the solvent with all of the solvent molecules expelled from the IS region by inclusion of the hard sphere in the sampling, $\beta\mu_{\mathrm{OS,LR}}^{\mathrm{ex}} (\lambda)$, where
\begin{equation}
\langle\mathrm{e}^{-\beta\Delta U } \rangle_{\lambda}
= \frac
{\int dx^N \mathrm{e}^{-\beta U_N}\mathrm{e}^{-\beta\Delta U_{\mathrm{HS}} (\lambda) } \mathrm{e}^{-\beta\Delta U  } }
{\int dx^N \mathrm{e}^{-\beta U_N}\mathrm{e}^{-\beta\Delta U_{\mathrm{HS}} (\lambda) } } .
\end{equation}
We can view this expression in a different way: 
\begin{eqnarray}
\langle\mathrm{e}^{-\beta\Delta U } \rangle_{\lambda} &
= & 
\frac
{\int dx^N \mathrm{e}^{-\beta U_N}\mathrm{e}^{-\beta\Delta U_{\mathrm{HS}} (\lambda) } \mathrm{e}^{-\beta\Delta U  } }
{\int dx^N \mathrm{e}^{-\beta U_N} }
\frac
{\int dx^N \mathrm{e}^{-\beta U_N} }
{\int dx^N \mathrm{e}^{-\beta U_N}\mathrm{e}^{-\beta\Delta U_{\mathrm{HS}} (\lambda) } } \notag  \\ 
 & = & 
 \frac
 {\langle \mathrm{e}^{-\beta\Delta U_{\mathrm{HS}} (\lambda)} \mathrm{e}^{-\beta\Delta U} \rangle_0}
 {\langle \mathrm{e}^{-\beta\Delta U_{\mathrm{HS}} (\lambda)} \rangle_0} \notag \\ 
 &=&
 \int d \epsilon \frac
{ P^{(0)} (\epsilon,r_{\mathrm{min}}>\lambda)}
{p_0 (\lambda)} 
 \mathrm{e}^{-\beta\epsilon} . 
 \label{condition}
\end{eqnarray}
$P^{(0)} (\epsilon,r_{\mathrm{min}}>\lambda)$ is the joint probability of observing the coupling 
energy $\epsilon$ and the condition $r_{\mathrm{min}}>\lambda$, where $r_{\mathrm{min}}$ is
the distance of closest approach of the solvent molecules to the solute.  But the ratio of probabilities in
Eq.~(\ref{condition}) is simply the conditional probability $P^{(0)} (\epsilon|r_{\mathrm{min}}>\lambda)$,
so 
\begin{equation}
\mathrm{e}^{-\beta\mu_{\mathrm{OS,LR}}^{\mathrm{ex}} (\lambda)} 
= \langle\mathrm{e}^{-\beta\Delta U } \rangle_{\lambda} 
= 
 \int d \epsilon 
 P^{(0)} (\epsilon|r_{\mathrm{min}}>\lambda)
 \mathrm{e}^{-\beta\epsilon} ,
 \label{condition1}
\end{equation}
where the label LR is for the long-ranged contribution to the free energy.
Thus, instead of inserting the HS solute in the sampling, we could sample the uncoupled system
and average the distribution of interaction energies 
conditioned on the absence of
solvent molecules in the IS region. 

We can re-express the OS long-ranged contribution in a form that includes the 
coupled sampling by a rearrangement:
\begin{eqnarray}
\langle\mathrm{e}^{-\beta\Delta U } \rangle_{\lambda} &
= &  \frac
{\int dx^N \mathrm{e}^{-\beta U_N}\mathrm{e}^{-\beta\Delta U_{\mathrm{HS}} (\lambda) } \mathrm{e}^{-\beta\Delta U  } }
{\int dx^N \mathrm{e}^{-\beta U_N}\mathrm{e}^{-\beta\Delta U_{\mathrm{HS}} (\lambda) } }  \notag \\
&= &  \frac
{\int dx^N \mathrm{e}^{-\beta U_N}\mathrm{e}^{-\beta\Delta U_{\mathrm{HS}} (\lambda) } \mathrm{e}^{-\beta\Delta U  } }
{\int dx^N \mathrm{e}^{-\beta U_N}\mathrm{e}^{-\beta\Delta U_{\mathrm{HS}} (\lambda) } 
\mathrm{e}^{-\beta\Delta U  }\mathrm{e}^{\beta\Delta U  } } \notag  \\
& = & \langle \mathrm{e}^{\beta\Delta U  } \rangle^{-1}_{\lambda + \Delta U} . 
\end{eqnarray}
The averaging in the last step occurs for a system containing a fully interacting solute particle in addition to a hard-core particle of size $\lambda$.  Then the excess chemical potential
in Eq.~(\ref{mu_QCT}) can also be written as
\begin{equation}
\beta\mu^{ex} = \ln x_0 (\lambda) - \ln p_0 (\lambda) +  \ln \langle \mathrm{e}^{\beta\Delta U}
 \rangle_{\lambda + \Delta U} .
\label{mu_QCT1}
\end{equation}
The averaging in the last term is equivalent to 
\begin{equation}
\mathrm{e}^{\beta\mu_{\mathrm{OS,LR}}^{\mathrm{ex}}(\lambda)} =
\langle\mathrm{e}^{\beta\Delta U } \rangle_{\lambda+\Delta U} 
= 
 \int d \epsilon 
 P (\epsilon|r_{\mathrm{min}}>\lambda)
 \mathrm{e}^{\beta\epsilon} .
 \label{condition2}
\end{equation}
We note that a relation similar to Eq.~(\ref{DOS4}) holds here with the distributions 
replaced by the conditioned ones, and the excess chemical potential replaced by the 
long-ranged contribution.
The approach of Eq.~(\ref{condition2}) has been taken by 
Pratt and coworkers.\cite{jshah07,dasth07,dast08}  It avoids issues of de-wetting/re-wetting which may occur if a hard particle reference
system is used, and the attractive interactions are subsequently turned on. The averaging that leads
to $P (\epsilon|r_{\mathrm{min}}>\lambda)$ involves the real solute in the sampling, and requires the 
observation of configurations in which no solvent molecules are located in the IS region.
For small cavity sizes, this approach is feasible, but observations 
of spontaneous larger cavities in water are rare. 

Shah, {\em et al.},\cite{jshah07} Asthagiri, {\em et al.},\cite{dasth07} and Asthagiri, {\em et al.}\cite{dast08} observed that the conditioning inherent in Eq.~(\ref{condition2}) 
leads to a near-Gaussian
form for the distribution of interaction energies.  This is because moving the solvent molecules
some distance from the solute leads to many similar interactions smaller in magnitude, and 
the interactions do not include 
large-deviation (repulsive) overlaps with the solute.
With no conditioning, the distributions tend more toward an extreme-value (exponential) form due to the overlaps
of the solute with nearby solvent molecules. 

So we can view the conditioning in one of two ways: either include the hard particle directly 
in the sampling, or average the distributions only over configurations that exhibit no 
penetration of the solvent molecules into the inner shell.   The two approaches are 
equivalent, but the former approach allows for more efficient sampling, especially if 
the IS radius becomes large.  In both cases, the effect of pushing the solvent molecules
out of the IS domain leads to the same, approximately Gaussian interaction energy distributions.  

Due to the near-Gaussian form for the distributions, a second-order cumulant expansion 
for the OS long-ranged contribution is appropriate. The expansion from Eq.~({\ref{mu_QCT})
yields 
\begin{equation}
\beta\mu_{\mathrm{OS,LR}}^{\mathrm{ex}} (\lambda) =  - \ln \langle \mathrm{e}^{-\beta\Delta U} \rangle_{\lambda} \approx 
\beta \langle \Delta U \rangle_{\lambda} 
- \frac{\beta^2}{2} \left[ \langle \Delta U^2 \rangle_{\lambda} - 
\langle \Delta U \rangle^2_{\lambda}  \right] + \ldots
\label{mu_QCTlr_cum}
\end{equation}
Similarly, Eq.~({\ref{mu_QCT1}) results in
\begin{equation}
\beta\mu_{\mathrm{OS,LR}}^{\mathrm{ex}} (\lambda) =   \ln \langle \mathrm{e}^{\beta\Delta U} 
\rangle_{\lambda+\Delta U} \approx 
\beta \langle \Delta U \rangle_{\lambda+\Delta U}  +
 \frac{\beta^2}{2} \left[ \langle \Delta U^2 \rangle_{\lambda+\Delta U} - 
\langle \Delta U \rangle^2_{\lambda+\Delta U}  \right] + \ldots
\label{mu_QCT1lr_cum}
\end{equation}
Thus the mean-field 
$ \langle \Delta U \rangle_{\lambda}$ and  $ \langle \Delta U \rangle_{\lambda+\Delta U}$
are upper and lower bounds, respectively, on the long-ranged part of the free energy by the Gibbs-Bogoliubov inequality.\cite{aisih68,verlet72,chbenn76,cchip07,ghum07,nlu07}  If the 
interaction energy distribution for either the coupled or the uncoupled parts is 
approximately Gaussian,
then the other is also near-Gaussian (with a similar width), and adding the two 
expressions leads to a near-cancellation of the fluctuation terms.  
Hence, a sensible approximation to test would
be 
\begin{equation}
\beta\mu_{\mathrm{OS,LR}}^{\mathrm{ex}} (\lambda)  \approx 
\frac{\beta}{2} \left[  \langle \Delta U \rangle_{\lambda}  +
\langle \Delta U \rangle_{\lambda+\Delta U} \right] .
 \label{mu_QCTlr_MFave}
\end{equation}
This formula is similar to one of the thermodynamic integration formulas developed
by Hummer and Szabo,\cite{ghumm961} but is obtained in a different context here.

\section{ Bayesian Estimates of Outer-Shell Packing and Inner-Shell Chemical Contributions}
\label{sec_Bayes}

In this section, we develop a Bayesian approach\cite{sivia,jaynes} for estimating both the 
$\mu_{\mathrm{OS,HS}}^{\mathrm{ex}} (\lambda) $ and $\mu_{\mathrm{IS}}^{\mathrm{ex}} (\lambda)$
contributions to the excess chemical potential.  This method generalizes a discussion of spatial stratification outlined in Ref.~\onlinecite{ourbook} (Ch.~5). 
The probabilities for the OS packing and IS chemical contributions,
Eqs.~(\ref{p_0}) and (\ref{x_0}), relate to the probability that a defined region of space is unoccupied, without or with the solute present, respectively.  The occupancy condition can be reduced to monitoring the distance from the solute 
to the center of the nearest solvent molecule $r_{\mathrm{min}}$.  This creates a one-to-one relationship between the probability distribution of the closest solvent molecule $\func{P}{}{r_{\text{min}}}$ and the probability that determines the above chemical potentials $\func{P}{}{r_{\text{min}} > \lambda_i}$:
\begin{equation}
p (\lambda_i) \equiv \func{P}{}{r_{\text{min}} > \lambda_i}
          =  \int^{\infty}_{\lambda_i}{\func{P}{}{r_{\text{min}}} dr_{\text{min}} }
	= 1-\int_0^{\lambda_i}{\func{P}{}{r_{\text{min}}} dr_{\text{min}} } .
\label{eq:probrmin}
\end{equation}
This means that calculating the chemical potential profile for different values of $\lambda_i$ reduces to a problem of accurately estimating the distribution of $r_{\text{min}}$.  

To exhibit the physical background of the above discussion, we first review the relation of the distribution of closest approach to the scaled particle theory (SPT) for the packing part of the free energy.  An integral equation for the distribution $\func{P}{}{r_{\text{min}}}$ can be derived by noting that this distribution is the probability of observing a cavity of size $r_{\text{min}}$ {\em and} a solvent particle in the range $r_{\text{min}} \rightarrow r_{\text{min}} + dr_{\text{min}}$ (see Fig.~\ref{fig:hertz}).  This is just the conditional probability of observing a solvent particle in the window given an $r_{\text{min}}$-sized cavity $\left( 4 \pi r_{\text{rmin}}^2 G(r_{\text{min}}) \rho \right)$ times the cavity probability $\left( \left[ 1-\int_0^{r_{\text{min}}}{\func{P}{}{s} }ds \right] \right)$, where $G(r_{\text{min}})$ is the contact value of the radial distribution function for the hard sphere interacting with the solvent, and $\rho$ is the solvent density:
\begin{equation} 
\func{P}{}{r_{\text{min}}} = 4 \pi r_{\text{min}}^2 \rho G(r_{\text{min}}) 
\left[ 1-\int_0^{r_{\text{min}}}{\func{P}{}{s} }ds \right] ,
\label{eq:hertz1}
\end{equation}
The above integral equation can be solved analytically, yielding
\begin{equation}
\func{P}{}{r_{\text{min}}} = 4 \pi r_{\text{rmin}}^2 \rho G(r_{\text{min}}) 
\exp \left( - \int_0^{r_{\text{min}}} 4 \pi s^2 \rho G(s) ds \right) .
\label{eq:hertz2}
\end{equation} 
When this expression is inserted into Eq.~(\ref{eq:probrmin}) for $\func{P}{}{r_{\text{min}} > \lambda_i}$, the resulting excess chemical potential is
\begin{equation}
\beta \mu_{\text{OS,HS}}^{\text ex} (\lambda) = \int_0^{\lambda} 4 \pi r^2 \rho G(r)  dr ,
\label{eq:hertz3}
\end{equation}
which is just the SPT expression.\cite{hsashb06}
If $G(r) = 1$ in the above equation (ideal gas case), we obtain the Hertz distribution for $\func{P}{}{r_{\text{min}}}$ with the excess chemical potential $\beta \mu_{\text{OS,HS}}^{\text ex} (\lambda) = 4\pi \lambda^3 \rho / 3$.\cite{schan43,ggraz03}  Instead of solving directly for the contact value $G(r)$, here we develop a Bayesian estimation approach for obtaining $p(\lambda_i)$ in Eq.~(\ref{eq:probrmin}) from occupancy statistics data (in discretized spatial shells) assembled during the simulations.

In order to calculate the OS packing and IS chemical potential contributions,
Eqs.~(\ref{p_0}) and (\ref{x_0}), we conduct two sets of simulations with the solute having either a zero interaction or a full interaction with the solvent, respectively.
Thus in this section we will make the assignment $\func{P}{}{r_{\mathrm{min}} > \lambda_i} \equiv p_i$, without reference to the particular choice of sampling (no solute, with solute, etc.).
It should be kept in mind that the $p_0$ term in the OS packing contribution $\beta\mu_{\mathrm{OS,HS}}^{\mathrm{ex}} (\lambda_i) = - \ln \func{P}{0}{r_{\mathrm{min}} > \lambda_i}$ is not the $i=0$ limit case of $p_i$.  Rather, $\func{P}{0}{r_{\mathrm{min}}>\lambda_i} = p_i$ for the case of no solute present in the sampling.  When calculating the IS probability, then $\func{P}{\Delta U}{r_{\mathrm{min}}>\lambda_i} = p_i$, and the solute is included in the sampling.

Assume we have collected data from a set of simulations including hard-core solutes of varying sizes,
$\lambda_j$, indexed in ascending order by $j = 0, 1, \ldots , L-1$, and we are interested in calculating the free energy
$\ln p_L$ for some arbitrary distance $\lambda_L > \lambda_{L-1}$. We discretize the problem by separating the possible values of $r_{\text{min}}$ into shells around the solute (or cavity center) as in Fig.~\ref{fig:schem}.
This creates a set of mutually independent and exhaustive events, whose probabilities are
\begin{equation}
s_k = \func{P}{}{r_{\text{min}} \in \left(\lambda_k,\lambda_{k+1}\right]} ,
\end{equation}
where the $\lambda$ values below $\lambda_L$ must coincide with the set of hard-core sizes included in the simulations.
The data gathered from the $j^{\text{th}}$ simulation (which included a hard-core particle of size $\lambda_j$) consists of accumulating the counts $x_j(k)$ of occupancy events for the nearest solvent molecule in the shells labeled by $k$.

The likelihood of such a set of observations from the $j^{\text{th}}$ hard-sphere simulation radius is the ratio of two multinomial functions:
\begin{align}
&\func{P}{}{\{x_j(k)\} | N_j, \{s_k\}, \lambda_j } \notag \\
&\qquad= \frac{\func{P}{}{\{x_j(k), k=j,j+1,\ldots,L\} | N_j, \{s_k\}}}
   {\func{P}{}{\{x_j(k)=0; k=0,1, \ldots,j-1\}| N_j, \{s_k\}}} \notag \\
&\qquad= \frac{N_j! \prod_{k=j}^L \left(s_k/p_j\right)^{x_j(k)}}
        {\prod_{k=j}^L x_j(k)!} \notag \\
&\qquad= A(x) p_j^{-N_j} s_L^{x_j(L)} \prod_{k=j}^{L-1} s_k^{x_j(k)}  ,
\end{align}
where the first equality expresses the conditional probability as the joint probability of the observed occupancies and an existing cavity divided by the probability of the observation of the cavity; the further equalities follow from inserting the multinomial expressions.  
Here $N_j$ is the total number of observations from the simulation and $A$ is some function of $x$ in which we are not interested, since our final goal is the posterior probability of $\{s_k\}$ (and thus $p_i$).

The prior chosen here is that of Haldane and Zellner, i.e.~an improper Dirichlet distribution with zero initial observations:\cite{mzhu04}
\begin{equation}
\func{P}{}{N_j, \{s_k\}, \lambda_j } \propto \prod_{k=0}^{L} s_k^{-1} .
\end{equation}
This noninformative prior is a commonly used one for the multinomial case; it emphasizes
the limits of the occupancy distributions, resulting in a rapid convergence to the final posterior distribution with increased sampling.\cite{sivia}
Other choices could have been made, such as incorporating smoothness in the $s_k$ profile.  The effect of the prior distribution diminishes, however, as more observations are gathered.\cite{sivia}

The total likelihood function is a product of likelihoods from all simulations.  The posterior distribution for the shell probabilities $\{s_k\}$ is calculated by multiplying the data likelihood with the prior distribution and normalizing:
\begin{equation}
\func{P}{}{\{s_k\} | N x} \propto s_L^{\zeta_L-1} \prod_{i=0}^{L-1} p_i^{-N_i} s_i^{\zeta_i-1} .
\end{equation}
This procedure gives a complicated function of shell probabilities owing to the division by $p_j = 1-\sum_{k=0}^{j-1} s_k$.
Since we are interested in deriving a posterior probability distribution that aids in estimating the logarithm of $p_L$, it is useful to carry out a transformation of variables from the shell probabilities to conditional increments $\left\{ \delta_i=\func{P}{}{r_{\text{min}} > \lambda_i | r_{\text{min}} > \lambda_{i-1}} \right\}$. These conditional probabilities are closely related to the intensity function of Gumbel,\cite{gumbel} and they refer to the probability of observing no solvent particles between $\lambda_{i-1}$ and $\lambda_i$, given that the region inside $\lambda_{i-1}$ is empty.  
In terms of these new variables, the logarithm of $p_L$ is additive (since $p_j = \prod_{i=1}^j \delta_i$).
To connect with the SPT discussion above, we note that $s_i = p_i (1-\delta_{i+1})$. The mapping here is that $s_i$ is the probability of closest approach, $p_i$ is the cavity probability, and $(1-\delta_{i+1})$ is the conditional probability of finding a solvent molecule in the shell given the existence of the cavity.  Thus we can note that the $4 \pi r_{\text{min}}^2$ appearing in Eq.~(\ref{eq:hertz1}) is carried by $(1-\delta_{i+1})$. Then the Jacobian for the transformation is $\prod_{i=1}^L \delta_i^{L-i}$, and the posterior becomes:
\begin{equation}
\func{P}{}{ \{\delta_i\} | N x} \propto p_L^{\zeta_L-1}
    \prod_{i=0}^{L-1} \delta_i^{L-i} p_i^{\zeta_i-N_i-1}
{\left(1-\delta_{i+1}\right)}^{\zeta_i-1} ,
\label{eq:newposterior}
\end{equation}
where we have defined bin totals as
\begin{align}
\zeta_k &= \sum_{j=0}^k x_j(k) \quad k=0,1,\ldots,L-1 \\
\zeta_L &= \sum_{j=0}^{L-1} x_j(L) = \sum_{j=0}^{L-1} N_j
                                    - \sum_{k=0}^{L-1} \zeta_k \notag .
\end{align}

This transformation of variables has thus led us to a simple form for the posterior distribution.  Because the posterior of Eq.~(\ref{eq:newposterior}) is a product and does not contain cross-terms between the $\delta_i$, it shows that our estimate for each $\delta_i$ follows an independent beta distribution 
\begin{equation}
\func{P}{}{\delta_i | N x} = \tfrac{\Gamma(\alpha_i+\beta_i)}
      {\Gamma{(\alpha_i)}\Gamma{(\beta_i)}} \delta_i^{\alpha_i-1}
(1-\delta_i)^{\beta_i-1} ,
\end{equation}
with parameters
\begin{align}
\label{eq:halpha}
 \alpha_i &= \sum_{j=0}^{i-1} N_j - \sum_{k=0}^{i-1} \zeta_k \\
\beta_i &= \zeta_{i-1} .
\end{align}

This is what we might have suspected from the start, since the number of counts in bin $k$ should really be combined from all simulations that can observe such an event.  The total number of observations used for determining each increment, ($\alpha_i + \beta_i$), is therefore all counts from these simulations ($\lambda_j < \lambda_i$) in shells $k \ge (i-1)$, i.e. a sub-block of Table~\ref{tbl:shells} including the lower left corner.  The above discussion also warns us that the probability for any increment will be more uncertain if either $ \alpha_i$ or $ \beta_i$ is small (corresponding to no counts above ${\lambda}_i$ or no counts inside 
$\left({\lambda}_{i-1},{\lambda}_i\right]$, respectively).

The free energy is constructed by adding together the incremental free energies:
\begin{equation}
\ln(p_L) = \sum_{i=1}^L \ln(\delta_i) .
\end{equation}
We derive the Bayesian minimum mean-squared error estimator\cite{jaynes}
by integrating over the posterior pdf:
\begin{align}
\func{E}{}{\ln \delta} &= \tfrac{\Gamma(\alpha+\beta)}
      {\Gamma{(\alpha)}\Gamma{(\beta)}}
\int{d\delta \text{ } \delta^{\alpha-1} (1-\delta)^{\beta-1} \ln \delta} \notag\\
&= \psi_0( \alpha) - \psi_0( \alpha +  \beta) .
\end{align}
Here $\psi_n(z)$ is the polygamma function.\cite{abramsteg} A similar integral 
can be derived for the variance (below).  Since the the means and variances of independent random variables are additive under summation, the Bayesian estimates for the free energy and its variance are then
\begin{align}
\func{E}{}{\ln(p_L)} &= \sum_{i=1}^L
       \psi_0( \alpha_i) - \psi_0( \alpha_i +  \beta_i) \\
\func{E}{}{\left(\ln(p_L)-\func{E}{}{\ln(p_L)}\right)^2} &= \sum_{i=1}^L
	\psi_1( \alpha_i) - \psi_1( \alpha_i +  \beta_i) .
\end{align}

The simplest way to carry out the whole calculation is to tabulate $x_j(k)$ as in Table~\ref{tbl:shells} and compute partial sums over sub-blocks to obtain a set of $L \text{ }  \alpha_i \text{ and }  \beta_i$ values.  This process must be repeated for each choice of ${\lambda}_L$ at which the inner-shell or outer-shell packing excess chemical potential value is to be estimated. We used the scipy special python library implementation of the polygamma function for the calculations in this paper.\cite{scipy}   At the end of the next section, we discuss how this Bayesian approach to occupancy statistics can be implemented using molecular dynamics simulation data.

\section{Inclusion of a continuous model potential for molecular dynamics sampling}
\label{sec_MD}

Our goal is to implement the approach outlined above in large-scale molecular dynamics simulations.   In order to carry out the indicated averages involving hard-sphere particles, we can employ a change of system energy function to one including a repulsive but continuous model potential, $M(r)$.  The model potential is chosen such that the hard-sphere condition is likely satisfied. A helpful expression results from the PDT formula for averages:\cite{ourbook}
\begin{equation}
\left\langle F \right\rangle_{\lambda+M} = 
\frac
{\left\langle \mathrm{e}^{-\beta\Delta U_{\mathrm{HS}} (\lambda)} F \right\rangle_M}
{\left\langle \mathrm{e}^{-\beta\Delta U_{\mathrm{HS}} (\lambda)}  \right\rangle_M} =
\left\langle F | r_{\mathrm{min}} > \lambda \right\rangle_M .
\end{equation}
With this approach the averaging with the hard-sphere and model potentials included is re-written as a conditional average with only the model potential in the sampling.  Since the repulsive model potential is chosen to closely mimic the hard-sphere potential, relatively few configurations are discarded, and good statistics can be obtained.  

We first focus on the long-ranged OS contribution to the excess chemical potential,
$\beta\mu_{\mathrm{OS,LR}}^{\mathrm{ex}}$.
We re-express that contribution as
\begin{align}
\mathrm{e}^{-\beta \mu^{\mathrm{ex} }_{\mathrm{OS,LR}}(\lambda)} &=
\frac{\int{ dx^N \mathrm{e}^{-\beta U_N} \mathrm{e}^{-\beta\Delta U_{\mathrm{HS}}(\lambda)}
 \mathrm{e}^{-\beta\Delta U} }}
     {\int{dx^N \mathrm{e}^{-\beta U_N} \mathrm{e}^{-\beta \Delta U_{\mathrm{HS}}(\lambda)} }} \notag \\ &=
\frac{{\left\langle \mathrm{e}^{-\beta\left[ \Delta U - M \right]} \right\rangle}_{\lambda+M}}
     {{\left\langle \mathrm{e}^{\beta M} \right\rangle }_{\lambda+M}}   \notag \\ &=
\frac {{\left\langle \mathrm{e}^{-\beta \left[ \Delta U - M \right]} \, | r_{\text{min}} > \lambda \right\rangle}_{M}}
     {{\left\langle \mathrm{e}^{\beta M} \, | r_{\text{min}} > \lambda \right\rangle }_M} ,
\label{eq:modellr1}
\end{align}
or in the inverse form
\begin{align}
\mathrm{e}^{\beta \mu^{\mathrm{ex} }_{\mathrm{OS,LR}}(\lambda)} &=
\frac{\int{ dx^N \mathrm{e}^{-\beta U_N} \mathrm{e}^{-\beta\Delta U_{\mathrm{HS}}(\lambda)}
 \mathrm{e}^{-\beta\Delta U}  \mathrm{e}^{\beta\Delta U}}}
     {\int{dx^N \mathrm{e}^{-\beta U_N} \mathrm{e}^{-\beta \Delta U_{\mathrm{HS}}(\lambda)} } 
      \mathrm{e}^{-\beta\Delta U} }\notag \\ &=
\frac{{\left\langle \mathrm{e}^{\beta\left[ \Delta U + M \right]} \right\rangle}_{\lambda+M+\Delta U}}
     {{\left\langle \mathrm{e}^{\beta M} \right\rangle }_{\lambda+M+\Delta U}}   \notag \\ &=
\frac {{\left\langle \mathrm{e}^{\beta \left[ \Delta U + M \right]} \, | r_{\text{min}} > \lambda 
\right\rangle}_{M+\Delta U}}
     {{\left\langle \mathrm{e}^{\beta M} \, | r_{\text{min}} > \lambda \right\rangle }_{M+\Delta U}} .
\label{eq:modellr2}
\end{align}

Cumulant expansions can be performed on Eqs.~(\ref{eq:modellr1}) and (\ref{eq:modellr2}), 
leading to the following relations (to second order in $\beta$):
\begin{eqnarray}
\label{eq:MDref}
\beta\mu^{\mathrm{ex}}_{\mathrm{OS,LR}} & \approx& \beta {\left\langle \Delta U | r_{\mathrm{min}}>\lambda \right\rangle}_{M} - \frac {\beta^2}{2} \left( V_{UU} + V_{MM} - 2 V_{UM} \right) + \frac{\beta^2}{2} V_{MM} \notag \\
&& \le \beta {\left\langle \Delta U| r_{\mathrm{min}}>\lambda \right\rangle}_{M} + \frac{\beta^2}{2} V_{MM}
\label{eq:OSLRapprox}
\end{eqnarray}
and
\begin{eqnarray}
\label{eq:MDfull}
\beta\mu^{\mathrm{ex}}_{\mathrm{OS,LR}} & \approx& \beta {\left\langle \Delta U | r_{\mathrm{min}}>\lambda \right\rangle}_{M+\Delta U} + \frac {\beta^2}{2} \left( W_{UU} + W_{MM} + 2 W_{UM} \right) - \frac{\beta^2}{2} W_{MM} \notag \\
&& \ge \beta {\left\langle \Delta U| r_{\mathrm{min}}>\lambda \right\rangle}_{M+\Delta U} - \frac{\beta^2}{2} W_{MM} .
\label{eq:OSLRapprox1}
\end{eqnarray}
Here $V_{UU}, V_{MM}, W_{UU}$ and $W_{MM}$ are variances:
\begin{eqnarray}
V_{UU} & = & \left\langle \Delta U^2 | r_{\mathrm{min}} > \lambda \right\rangle_M
 - \left\langle \Delta U | r_{\mathrm{min}} > \lambda \right\rangle_M^2 \\
 V_{MM} & = & \left\langle M^2 | r_{\mathrm{min}} > \lambda \right\rangle_M
 - \left\langle M | r_{\mathrm{min}} > \lambda \right\rangle_M^2 \\
 W_{UU} & = & \left\langle \Delta U^2 | r_{\mathrm{min}} > \lambda \right\rangle_{M+\Delta U}
 - \left\langle \Delta U | r_{\mathrm{min}} > \lambda \right\rangle_{M+\Delta U}^2 \\
 W_{MM} & = & \left\langle M^2 | r_{\mathrm{min}} > \lambda \right\rangle_{M+\Delta U}
 - \left\langle M | r_{\mathrm{min}} > \lambda \right\rangle_{M+\Delta U}^2 , 
\end{eqnarray}
 and $V_{UM}$ and $W_{UM}$ are covariances:
\begin{eqnarray}
V_{UM} & = & \left\langle \Delta U M | r_{\mathrm{min}} > \lambda \right\rangle_M
 - \left\langle \Delta U | r_{\mathrm{min}} > \lambda \right\rangle_M
 \left\langle M | r_{\mathrm{min}} > \lambda \right\rangle_M \\
  W_{UM} & = & \left\langle \Delta U M | r_{\mathrm{min}} > \lambda \right\rangle_{M+\Delta U}
 - \left\langle \Delta U | r_{\mathrm{min}} > \lambda \right\rangle_{M+\Delta U}
 \left\langle M | r_{\mathrm{min}} > \lambda \right\rangle_{M+\Delta U} .
\end{eqnarray}
In this way, bounds can still be recovered from the model potential simulations.  For the rest of the discussion, the $M$-sampled system will be referred to as the {\em reference} system, and the $(M+\Delta U)$-sampled system will be referred to as the {\em coupled} system; the {\em fully coupled} system is recovered when no hard-sphere condition or model potential is applied in a coupled simulation.

The conditional variances of the model potential ($V_{MM}$ and $W_{MM}$) appear as extra quantities widening the bounds.  The best choice of model potential would therefore minimize these variances as much as possible.  To accomplish this, note that for any model potential that is zero outside the chosen hard sphere radius,  $V_{MM}$ and $W_{MM}$ become identically zero.  On the other hand, if the model potential has a repulsive wall well inside the hard sphere radius, many configurations will violate the hard-sphere condition and the simulations will not yield good statistics.  Since the aim of introducing the model potential is to allow molecular dynamics sampling, it must be continuous and have a derivative that does not make the system numerically unstable with the chosen time-step.  These considerations suggest choosing $M(r)$ to be as steep as possible at the chosen hard-sphere boundary.  

For the calculations performed here, a Lennard-Jones potential with Weeks-Chandler-Anderson (WCA)\cite{wca} style truncation at the minimum was used for $M(r)$.  The coefficients were chosen to make the model potential's value at the hard sphere boundary equal to a defined constant $E_\lambda$ and to force the minimum to coincide with a given cutoff $R_{c,\lambda} > \lambda$:
\begin{align}
M(r) &= \begin{cases}
c_{12}/r^{12} - c_6/r^6 + c_{12}/R_{c,\lambda}^{12} & r \leq R_{c,\lambda} \notag \\
  0 & r > R_{c,\lambda}
\end{cases} \\
c_6 &= {\frac{2 c_{12}}{R_{c,\lambda}^6 }}, \quad
c_{12} = \frac {E_\lambda} {\left( \lambda^{-6} - R_{c,\lambda}^{-6}   \right)^2} .
\end{align}
Initial simulations used a value of $k_B T/2$ (where $k_B$ is the Boltzmann constant and $T$ is the temperature) for $E_\lambda$, and the choice $R_{c,\lambda}=\lambda+0.1${\AA} for the cutoff radius.  We noticed, however, that more configurations were rejected at larger model potential radii and from the coupled systems due to violations of the hard sphere conditions, giving worse sampling for these cases.  
Various alterations of the model potential were explored to enhance the number of accepted configurations.  For coupled simulations of Na$^+$ and Cl$^-$ and large-$\lambda$ simulations of CH$_4$ and CF$_4$, we used a new choice for the cutoff radius, $R_{c,\lambda}=1.05 \lambda$, but the ion simulations continued to exhibit low acceptance probabilities. 

We then pursued an alternative strategy of setting the $M(r)$ parameters to fixed values and altering the HS size $\lambda$ downward, based on the distributions of solvent closest approach, until substantial acceptance probabilities (on the order of 80\%) were obtained. Similar alterations were also applied for the largest 
$\lambda$ simulations of TIP3P water.  A sequence of $M(r)$ potentials with increasing effective particle size, followed by adjustment of the HS size $\lambda$ to yield sizeable  acceptance probabilities, thus appears a sensible procedure for generating the data for a range of $\lambda$'s.  

The origin of the problem of determining the $M(r)$ form is that the accessible volume for penetration of solvent molecules inside the specified hard-sphere 
$\lambda$ increases for larger radii.  Future work should explore alternative model potentials that are harsher near the $\lambda$ radius.  We note, however, that the size range considered in our simulations (below) is already quite large for the cases of interest, and the hard cores of larger molecules can be considered as additive contributions from multiple hard spheres with typical sizes on the order of those examined here.     

In the present study, free energy bounds on the OS long-ranged part of the free energy were obtained as discussed above.  In order to obtain an accurate estimate of 
$\mu^{\mathrm{ex}}_{\mathrm{OS,LR}}$ based on 
Eq.~(\ref{mu_QCTlr_MFave}), a re-weighting strategy was employed, in which the data from
the model potential simulation yields an estimate of the mean-field energy with the HS particle
included in the sampling: 
\begin{equation}
{\left\langle\Delta U\right\rangle}_\lambda = \frac{
  {\left\langle \Delta U
     \mathrm{e}^{\beta M}|r_{\mathrm{min}} > \lambda \right\rangle}_M }
{ {\left\langle
     \mathrm{e}^{\beta M}|r_{\mathrm{min}} > \lambda \right\rangle}_M } .
\label{MHSrewt}
\end{equation}
A similar equation holds for the coupled case.  
 This re-weighting procedure is valid when the model potential is chosen small enough 
 outside $\lambda$ that adequate sampling of this area of configuration space is achieved.

Re-weighting is also necessary for calculating the shell occupancy observation numbers $x_j(k)$ (for the Bayesian analysis of Section \ref{sec_Bayes})
from model potential simulations.  Assuming $N_j$ independent samples 
satifsying the HS condition have been collected from a model potential simulation mimicking a hard sphere of radius $R_j$, the reweighting is given by
\begin{equation}
x_j(k) = N_j \frac{\left\langle \mathrm{e}^{\beta M_j}
	| r_{\mathrm{min}} \in \left( {\lambda}_k, {\lambda}_{k+1} \right] \right\rangle}
			{\left\langle e^{\beta M_j}
	| r_{\mathrm{min}} > {\lambda}_j \right\rangle},\quad k \ge j .
\end{equation}

As a check on our mean-field average formula, Eq.~(\ref{mu_QCTlr_MFave}),
the histogram overlap method can be applied to straight-forwardly  estimate 
$\beta \mu^{\mathrm{ex}}_{\mathrm{OS,LR}}$: 
 \begin{equation}
\beta \mu^{\mathrm{ex}}_{\mathrm{OS,LR}}
= \beta \Delta U + \ln { \frac
{P_{M+\Delta U}(\Delta U | r_{\mathrm{min}} > \lambda)}
{P_{M}(\Delta U | r_{\mathrm{min}} > \lambda)} }
 + \ln {\frac
{{\left\langle \mathrm{e}^{\beta M} \, | r_{\mathrm{min}} > \lambda \right\rangle}_M}
{{\left\langle \mathrm{e}^{\beta M} \, | r_{\mathrm{min}} > \lambda \right\rangle}_{M+\Delta U}}} .
\end{equation}
A direct approach involves employing the BAR method\cite{chbenn76} on the interaction energy distributions from the model potential simulations.
The BAR estimation is enacted on the first two terms on the rhs above.
The last term serves as the constant correction to the BAR estimate of the free energy difference between coupled and reference model particle simulations.  We utilized this approach to calculate $\beta \mu^{\mathrm{ex}}_{\mathrm{OS,LR}}$ for TIP3P water in order to judge the accuracy of the Gaussian approximation, Eq.~(\ref{mu_QCTlr_MFave}).  This approach can only be employed, of course, when some overlap exists between the interaction energy distributions.

To conclude this section, we outline a strategy for implementing the above theory in molecular dynamics simulations.   We first perform simulations with the solute included and then with no solute (and no model potential).  The occupancy counts for the distance of closest approach of the solvent to the solute in the first case, and to the cavity center in the second case, are assembled. The quantities $x_0 (\lambda)$ and $p_0 (\lambda)$ can then be computed out to radii for which good statistics are obtained in those probability distributions.  The degradation of the statistics at larger radii is easily monitored by plotting the estimated variances in addition to the chemical potentials.  At the radii where the variances become large, model potentials $M(r)$ are inserted to begin to force solvent molecules further from the solute.  Occupancy observation numbers  $x_j(k)$ are assembled in each of the simulations for the subsequent Bayesian analysis leading to $x_0 (\lambda)$ and $p_0 (\lambda)$.  This process is continued until radii are reached at which the Gaussian approximation for the long-ranged contribution becomes accurate.  

The same set of simulations of the reference and coupled systems yield estimates of the OS long-ranged part of the free energy $\beta \mu^{\mathrm{ex}}_{\mathrm{OS,LR}}$. The two terms in Eq.~(\ref{mu_QCTlr_MFave}) are estimated 
via Eq.~(\ref{MHSrewt}) and its coupled simulation counterpart.  A check of the mean-field approximation Eq.~(\ref{mu_QCTlr_MFave}) can be obtained by computing the total fluctuation terms in the first expressions of Eqs.~(\ref{eq:OSLRapprox}) and (\ref{eq:OSLRapprox1}); the extent of cancellation of these fluctuations monitors how closely the energy distributions mimic Gaussian behavior (more appropriately, re-weighted forms of the flucutations can be computed as was done for the mean-field term in Eq.~(\ref{MHSrewt}), see below).  Alternatively, the long-ranged contribution can be obtained by the BAR method discussed above, and the accuracy of the mean-field approach can be assessed directly for cases where some overlap of the energy distributions for the reference and coupled simulations occurs.  Typically, only a few model potential simulations need to be performed during the `pushing out' process in order for this procedure to converge to results close to the exact free energy for the cases examined here.  Numerical tests of this procedure are presented below.

\section{Molecular dynamics simulations}
\label{sec_Simulations}

In order to test the formalism, several solutes 
ranging from nonpolar to polar to ionic were examined. The
CH$_4$ and CF$_4$ molecules and the  Na$^+$ and Cl$^-$ ions were each simulated in the NPT ensemble using 
the Gromacs version 3.3.1 code.\cite{elind01}  The solutes were modeled in solvent boxes that included
516 SPC/E waters.  The Lennard-Jones potential parameters for the various molecules and ions simulated are given in Table \ref{tbl:ffparam}, and the particle-mesh Ewald (PME) 
method\cite{uessm95} was employed to handle the long-ranged electrostatics interactions.   For reference, the solute/solvent radial distribution functions for each of the above solutes and for the TIP3P water case (below) are presented in Fig.~\ref{fig:RDFplots}.

The temperature was taken as 298.15 K and the pressure as 1 bar.  Fully coupled systems were equilibrated for 2 ns with the Berendsen temperature and pressure-coupling algorithms. Then 2 frames per picosecond were recorded during 5 ns of production runs that employed the Nose-Hoover thermostat and the Parrinello-Rahman barostat.  The OS long-ranged free energy contributions were calculated using two simulations corresponding to the two sampling distributions indicated in Eqs.~\ref{eq:MDref} and \ref{eq:MDfull} for each value of the solute-to-water oxygen exclusion distance $\lambda$.  Simulations including repulsive model potentials were run starting from the end of the fully coupled simulations.  Nonpolar solutes were equilibrated for 1 ns, while ions were equilibrated for 0.5 ns followed by production runs of 5 ns (nonpolar) or 1 ns (ions).  A cubic nonbonded interaction switching function was employed for the Lennard-Jones interactions between 10 and 11 {\AA}, and no cutoff correction was used for these interactions.

We also computed the solvation free energy of TIP3P water\cite{wjorg83} in water.  These simulations were conducted since the TIP3P model is widely used in biophysical simulations, and previous results have been obtained for comparison.  The TIP3P water simulations were carried out similar to the above, with 216 waters and a 9 {\AA} Lennard-Jones cutoff.  Production runs were of 4.5 ns duration following equilibration periods of 0.7 ns.  Comparison results can be found in Refs.~\onlinecite{jherm88} (-22.6 kJ/mol) and \onlinecite{rwhit04} (-23.43 kJ/mol).  Although these previous studies did not use Ewald summation, a study on a different water model \cite{jquin92} showed little difference between cut-off based and Ewald-based electrostatic treatment for the solvation free energy calculations on water. We note that the free energy results obtained by Paliwal {\em et al.}\cite{apali06} differ somewhat since a slightly different TIP3P water model\cite{ener96} was employed in that work.  

It is important to note that the difference in electrostatic energy between the charged (coupled) and uncharged (reference) states necessary to calculate the long-ranged free energy contribution for ions includes a net total charge correction defined in Ref.~\onlinecite{ourbook}, Eq.~5.50.   Gromacs version 3.3.1\cite{elind01} includes this correction, so the total electrostatic energy difference can be calculated from a simple re-run through the trajectory at a different ion charge state.

\section{Results}
\label{sec_Results}

We begin by discussing the free energy results as a function of the IS/OS partitioning radius for the various molecular and ionic cases considered here.  Due to the relatively small size of the systems examined, extensive statistical sampling was performed to obtained fully converged results.  Below we will examine the issues of convergence rates with simulation time and the number of required model potential simulations for statistically reliable results.  The main focus of these initial free energy results is to assess the accuracy of the QCT free energy calculation scheme presented above.

We start with the OS packing contribution, $\mu_{\mathrm{OS,HS}}^{\mathrm{ex}}(\lambda)$, whose calculation does not depend on the solute-solvent interaction energy.
That packing contribution was computed from simulations of pure water and the reference (model) potential systems used in this study; the results are presented in Fig.~\ref{fig:FEplots}. The occupancy data was processed using the Bayesian approach of Section \ref{sec_Bayes}. The model potential $M(r)$ was identical for the CH$_4$, CF$_4$, Na$^+$, and Cl$^-$ systems due to use of the same water model and excluded volume definition.  TIP3P water requires its own free energy  calculation, and the results for this model are increasingly lower than the SPC/E results for larger $\lambda$ values (off by 3.7 kJ/mol at 4 {\AA}). This discrepancy is mostly due to the slightly lower density of TIP3P water at 1 bar.  In line with the Gaussian information theory estimate,\cite{ghumm96} we found that the packing contribution can be estimated fairly well from the mean and variance of the occupancy number distribution up to $\lambda$ values of around 3.3 {\AA}.  This method overestimates the solvation free energies for larger $\lambda$ values, however, due to cooperative de-wetting transitions near a cavity.\cite{hsashb06}  The exact results start at zero for $\lambda = 0$ and increase to large positive values for large radii.

Another observation is that the revised scaled-particle theory (SPT) \cite{hsashb06} expression
\begin{equation}
\label{eq:SPT}
 \mu^{ex}_{HS} = -\frac{4 \pi k_B T \rho_W \lambda}{R} + \epsilon - 16 \pi R \gamma_{\infty} \delta
                      + 4 \pi R^2 \gamma_{\infty} + \frac{4 \pi}{3} R^3 p_{\text{sat}}
\end{equation}
accounts almost perfectly for the large-radius solvation behavior of our hard-core solutes, but must be carefully parameterized.  This theory attempts to model the transition between the microscopic  and macroscopic limits by least-squares fitting of the above 
$\mu_{\mathrm{OS,HS}}^{\mathrm{ex}}(\lambda)$ estimates to the SPT model.  The SPT model was calculated as a cubic interpolation of the microscopic (simulated) and macroscopic (above) expressions between two radii with $\gamma_{\infty}$ = 71.99 mN/m and neglecting the vapor saturation pressure.  Using calculated chemical potentials with radii spaced every 0.1 {\AA} from 2 to 3 {\AA} gave the best fit, with $\delta$ = 0.51 {\AA}, $\lambda$ = -3.95, and $\epsilon$ = -1.055.  Other ranges (2.5-3.5 and 3.0-4.0 {\AA}) gave positive $\delta$ and $\lambda$ parameters.  The Gaussian model and the revised SPT theory provide alternate simpler routes to the estimation of $\mu_{\mathrm{OS,HS}}^{\mathrm{ex}}(\lambda)$ where appropriate.  We note that our numerically exact results can be calculated out to quite large radii  due to the inclusion of the model potentials in the simulations.

The IS chemical potentials were computed for each test solute via the Bayesian method of Section 
\ref{sec_Bayes}; the results are presented in Fig.~\ref{fig:FEplots}.  Because this quantity is based on the distribution of the closest solvent atom, the same calculation method as that for the packing term was applied.  The statistical convergence and accuracy obtained here for the IS free energies results from the use of coupled simulation data at multiple $\lambda_i$ values.  The free energy for the IS term starts at zero for $\lambda = 0$, remains at zero until nonzero values in the solute/solvent radial distribution function are observed, and then steadily decreases to more negative values.   An interesting observation is that the IS free energies appear to decrease nearly linearly at the larger $\lambda$ values studied.  
This corresponds to an exponential decay in
$\func{P}{}{r_{\text{min}}>\lambda}$
for large $\lambda$.

Since the IS and OS packing contributions start at zero for small radii and move in opposite directions with increasing radius,
these two contributions tend to cancel to some extent.  
The sum of the IS and OS packing terms can be thought of as the work to carve out a cavity of size $\lambda$ followed by the free energy change to then release the nearby waters to make direct contact with the solute, so some extent of cancellation of these terms is to be expected.
For the nonpolar solutes, the magnitude of the packing term is larger than that of the IS term for all $\lambda$ values. The water case is interesting in that the OS packing and IS terms cancel at  $\lambda = 3.2${\AA} (where the OS packing excess chemical potential is 21.49 kJ/mol);  thus the OS long-ranged contribution yields the full solvation free energy at that radius.  Similarly, the Na$^+$ and Cl$^-$ ion cases display the same cancellation at 2.26 and 3.39 {\AA}, respectively.   This behavior is some indication of the differences between nonpolar and polar solvation: nonpolar solvation is dominated by packing contributions, while for polar solvation the local binding of solvent molecules to the solute molecule overcomes the packing penalty.

The remaining part of the present free energy partitioning scheme is the long-ranged contribution 
$\mu^{\mathrm{ex}}_{\mathrm{OS,LR}}(\lambda)$.  As discussed above, the conditioning inherent in this term (due to the `pushing out' of solvent molecules from direct contact with the solute) leads to near-Gaussian behavior in the interaction energy distributions, thus suggesting a second-order perturbation approach may be accurate.   We tested that proposal by computing the bounds on $\mu^{\mathrm{ex}}_{\mathrm{OS,LR}}(\lambda)$, and the approximate mean value using Eq.~(\ref{mu_QCTlr_MFave}). For the two nonpolar cases (CH$_4$ and CF$_4$) the bounds are very close to one another over the whole range of $\lambda$ values (Fig.~\ref{fig:FEplots}).  The OS long-ranged free energy contribution $\mu^{\mathrm{ex}}_{\mathrm{OS,LR}}(\lambda)$ first decreases with increasing $\lambda$, exhibits a minimum, and then begins to rise; the value must tend towards zero at large radii, but this asymptotic approach is very slow, even for the nonpolar solutes.  The initial decrease is due to the removal of any near hard-core overlaps in the sampling.  The free energy bounds are apparent for the smallest radii considered for the CF$_4$ molecule, indicating the fluctuation terms make a visible contribution for those small radii.  Nevertheless, the narrowness of the bounds is an indication of why first-order perturbation theory or the van der Waals approximation \cite{ourbook} works well for solutes of this type.  The total solvation free energy computed by summing the IS, OS packing, and long-ranged components from our Gaussian estimate is essentially exact for all radii considered.  These results clearly illustrate the lack of dependence of the QCT free energy on the conditioning radius $\lambda$.\cite{apali06}  

The inclusion of charged interactions for the water and ion cases results in a large increase in the separation of the upper and lower bounds for $\mu^{\mathrm{ex}}_{\mathrm{OS,LR}}(\lambda)$.  For  water, Fig.~\ref{fig:FEplots} shows that, for small exclusion radii, the separation of the bounds is nearly 80 kJ/mol, and decreases in magnitude to about 20 kJ/mol for the largest $\lambda$ values considered.  Comparison with the numerically exact BAR results shows that the mean-field average formula Eq.~(\ref{mu_QCTlr_MFave}) incurs significant errors up to an exclusion radius of roughly 3.1 {\AA}; for larger radii, the Gaussian approximation is nearly exact.  The sum of all three components of the free energy is correspondingly accurate beyond an exclusion radius of 3.1 {\AA}.   An indication of the error in the Gaussian approximation for the water case is also displayed in Table \ref{tbl:mferr}.   There we list the re-weighted fluctuation parts of the average of the two Gaussian approximations in Eqs.~(\ref{mu_QCTlr_cum}) and (\ref{mu_QCT1lr_cum}).  These fluctuation differences decrease with increasing exclusion radius, indicating the excellent accuracy of the Gaussian approximation.  In addition, it is apparent in Table \ref{tbl:mferr} that the bound-widening terms $V_{MM}$ and $W_{MM}$ are very small in magnitude for the model potentials used here.   

The impact of the conditioning on the water interaction energy distributions for the long-ranged contribution is shown in Fig.~\ref{fig:overlap}.  With no conditioning, the distributions are highly non-Gaussian and the interacting and non-interacting cases are widely different.  With increasing exclusion radius $\lambda$ in the sampling, the distributions become increasingly Gaussian with small deviations in the tails.  This is the physical justification of our mean-field average formula Eq.~(\ref{mu_QCTlr_MFave}).
 
Last, we discuss the results for the Na$^+$ and Cl$^-$ ions (Fig.~\ref{fig:FEplots}).  For these charged cases, no overlap of the interaction energy distributions occurs for any exclusion radius examined, and the free energy bounds are separated by $\approx$700 kJ/mol for the Na$^+$ case and $\approx$600 kJ/mol for the Cl$^-$ case.  Yet a small amount of conditioning results in accurate estimates of the total free energy, to within several kJ/mol compared with the numerically exact result; the errors range from 1.8 to 5.2 kJ/mol for the Na$^+$ ion and 0.82 to 7.8 kJ/mol for the Cl$^-$ ion, and thus are all within 2 kcal/mol of the exact result.  The largest contribution to the total free energy is the long-ranged component $\mu^{\mathrm{ex}}_{\mathrm{OS,LR}}(\lambda)$, comprising 95-99\% of the total at the chosen radii.  These results are remarkable in that two completely non-overlapping limits can be used to obtain an accurate free energy estimate with no intermediate sampling. This results from the conditioning due to the repulsive particle included in the sampling.   

To give some indication of the efficiency of the spatial conditioning approach described here for the IS and OS packing terms, we plot $\mu^{\mathrm{ex}}_{\mathrm{OS,LR}}(\lambda)$ and  $\mu^{\mathrm{ex}}_{\mathrm{IS}}(\lambda)$ as a function of $\lambda$, along with the mean plus and minus the variance for simulations for the TIP3P water case.   Fig.~\ref{fig:build} displays the results for a 4.5 ns simulation.  From this plot it is clear that good estimates of the IS and OS packing contributions can be obtained out to a $\lambda$ value of roughly 3.2 {\AA} using a simulation with no model potential.  Inclusion of two additional model potential simulations allows for accurate estimation of these contributions out to $\lambda = 3.5${\AA}, near the first minimum in the water-water $g(r)$.  In Fig.~\ref{fig:build}, we plot the corresponding data  for a 200 ps simulation.  It is clear that the reduced sampling leads to a larger variance at smaller radius, and two or three simulations are necessary to push accurate results out to a radius of about 3.3 {\AA}.  From simulations of other lengths (100 ps, 500 ps), it is apparent that increased simulation time leads to a decrease of the variance at a given radius.  From these results, we conclude that an accurate estimate of the IS and OS packing contributions for water can be obtained from a few model potential simulations over relatively short simulation times (roughly 600 ps total).  

For radius values in the range 3.1{\AA}$ < \lambda <$ 3.5{\AA}, the mean-field approximation for the long-ranged component discussed in this paper is quite accurate.  We also computed this quantity for each simulation time to test for convergence in the mean-field estimate of the long-ranged part of the free energy (water case). For a $\lambda$ value of 3.2 {\AA}, in a 200 ps simulation the OS long-ranged free energy contribution is converged to within 0.56 kJ/mol, and for a radius of 3.5 {\AA} the free energy is converged to within 0.79 kJ/mol relative to the 4.5 ns result.  Thus we can see that the long-ranged portion of the free energy is well-converged during the simulation times required to estimate the IS and packing parts of the free energy.

\section{Conclusions and Discussion}
\label{sec_Disc}

This paper presents a spatial stratification scheme for accurate and efficient estimation of absolute solvation free energies.   Quasi-chemical theory leads to a natural division of the solvation free energy into inner-shell (IS) and outer-shell (OS) components.\cite{ourbook,lpratt07}  The OS free energy can be further divided into packing and long-ranged contributions.  In this paper, an efficient Bayesian approach was developed to estimate the IS and OS packing contributions based entirely on occupancy statistics within shells specified by a hard-sphere cavity size $\lambda$.  A method for sampling the occupancy statistics during molecular dynamics simulations allows for efficient generation of the free energy profiles as a function of $\lambda$ out to large cavity radii.   Our spatial partitioning method differs from alternative stratification methods, such as those that soften the repulsive portion of the intermolecular potential\cite{tbeut94} for direct TI implementations; the present approach provides a more direct handling of the packing issues involved in absolute solvation free energy calculations.  

The long-ranged component of the OS free energy is the free energy of solvation of the solute with solvent excluded from the region specified by $\lambda$.  Thus this free energy results from sampling many configurations with energy contributions that have been conditioned to disallow penetration of the solvent into direct contact with the solute.  The distributions of the interaction energies then assume a near-Gaussian form owing to the combined contributions of many nearly independent, comparable-magnitude terms.\cite{dasth07}
This observation suggests that a second-order expansion of the free energy is accurate for this term; that expectation is born out in the computational results for a range of molecular and ionic solute types.  

The free energy calculations reported here require two sets of simulations that include continuous repulsive model potentials  during the sampling.  In the first (reference) set , the model potential particle is simulated while embedded in the solvent; in the second (coupled) set, the sampling includes the model potential particle with the full solute/solvent interaction potential included in addition.   Each of the two sets requires only a few simulations to generate the free energy profiles as a function of $\lambda$ with high accuracy.  The same sets of simulations can be used to generate the long-ranged contribution to the free energy.   The accuracy of that contribution can be monitored by examination of the cancellation of the fluctuation contributions to the free energy; once the reference and coupled estimates of the fluctuations nearly cancel, the mean-field estimate is of high accuracy.  Since this free energy estimate is the average of two mean-field terms, the calculations converge rapidly with simulation time.  The estimates are robust because each part of the calculation includes an accurate error estimate in addition to the mean value.

Several solutes were examined in this study, ranging in type from nonpolar to polar molecules and then ions.  The nonpolar solute cases exhibit narrow bounds on the long-ranged contribution, consistent with the fact that a van der Waals approach is appropriate.\cite{ourbook}  The bounds on the OS long-ranged free energy contribution are much wider for polar molecules and ions, yet with a relatively small degree of conditioning due to inclusion of a repulsive model potential, the mean-field estimates are quite accurate for the cases examined.  For the examples of the Na$^+$ and Cl$^-$ ions, the distributions of interaction energies are completely non-overlapping during the reference and coupled system sampling. Yet the spatial conditioning still yields reliable free energies with only the two limits sampled, due to the induced near-Gaussian energy distributions.  

The standard particle insertion approach to the PDT excess chemical potential\cite{dfren96} relies on the availability of accessible cavities in the solvent in which the solute samples favorable solvation environments.  Even though this approach is formally exact,  solutes of appreciable size are not likely to encounter instantaneous cavities of sufficient extent during standard molecular dynamics sampling of the solvent.  This provides a central motivation to the development of the quasi-chemical theory (QCT).   An example case where the QCT approach should find useful applications is in computing free energies for ions or molecules in highly restricted environments;  ion channels provide one such example.\cite{jyin04}  Another example might be water molecules in specific localized binding sites inside proteins.\cite{ghum08}  The approach developed here and elsewhere will allow for accurate pointwise absolute free energy estimates at specific locations in inhomogenous systems.

The long-term goal of the methods developed here is to examine specific ion effects related to Hofmeister series that impact on a wide range of chemical and biological phenomena.\cite{wkunz04}   Specific ion effects are often ascribed to one of two causes, ion size\cite{begg08} and polarizability.\cite{wkunz04} The effects tend to be larger for anions due to their higher polarizability and larger size relative to cations.   The ion size influence on free energies can be relatively easily estimated by focusing on the packing and long-ranged contributions in the QCT.   

The effects of anion polarizability (and solvent polarizability) are more challenging.  We suggest that the OS packing and long-ranged contributions are likely to be relatively accurately estimated by classical simulations, since for both these cases, the solvent is excluded from direct contact with the solute.  The IS component is the most-likely candidate for requiring an accurate quantum mechanical treatment, as has been discussed in a wide range of work\cite{ourbook} that recasts the IS term directly into a chemical equilibrium expression.   Limitations of that approach for modeling anion free energies have been noted,\cite{lpratt07} however, due to the large number of possible ion/water complexes with comparable energies. In anion/water cluster calculations, surface solvation states have been observed in addition to a large number of near-lying isomers.\cite{lper91,pjung02}  The solvation environment in the condensed phase can be expected to differ from that in clusters.\cite{srau02,jheu03,mmasia08}   

The approach developed in this paper provides a direct route to the IS part of the solvation free energy.  The only information required to compute that contribution is the solvent occupancy statistics within the region specified by the exclusion radius $\lambda$, and no computation of interaction energies is required.  Thus we suggest that this approach is well-suited for {\em ab initio} modeling of anion solvation free energies in the condensed phase.   Future work will attempt to extend the ideas developed here to those systems, with an aim of unraveling the various contributions to specific ion effects in solvation and ion binding to proteins.

\section{Acknowledgments}
We would like to gratefully acknowledge the support of the NSF (CHE-0709560), the Army MURI program (DAAD19-02-1-0227), and the DOE Computational Science Graduate Fellowship (DE-FG02-97ER25308) for the support of this work. We acknowledge the Ohio Supercomputer Center for a grant of supercomputer time.  We especially thank Lawrence Pratt for many helpful discussions.

\bibliography{bounds}

\newpage
\begin{table} 
  \caption{\label{tbl:shells}Shell occupancy data, $x_j(k)$, for calculating $\mu^{\mathrm{ex}}_{\mathrm{OS,HS}}$ in SPC/E water. The hard-sphere conditioning radius indexed by $j$ runs along the columns, and the shell index $k$ (from zero to 3.5 {\AA}) runs along the rows.}
  \begin{ruledtabular}
    \begin{tabular}{lrrrr}
 & ${\lambda}_0$\footnotemark[1] & $\lambda_1$ & $\lambda_2$ & $\lambda_3$ \\
\hline
$\left(0, {\lambda}_1\right]$ & 719415 & & & \\
$\left({\lambda}_1, {\lambda}_2\right]$ & 3607138 & 8052 & & \\
$\left({\lambda}_2, {\lambda}_3\right]$ & 831339 & 1821 & 8934 & \\
$\left({\lambda}_3, {\lambda}_{3.5} \right]$ & 2602 & 5 & 33 & 7438 \\
$\left({\lambda}_{3.5}, \infty \right)$ & 22 & 0 & 1 & 64 \\
    \end{tabular}
  \end{ruledtabular}
\footnotetext[1]{For ease of exposition, the subscripts correspond to the chosen shell radii (in {\AA}).}
\end{table}
.

\newpage
\begin{table*}
  \caption{Lennard-Jones force-field parameters}
  \label{tbl:ffparam}
  \begin{ruledtabular}
    \begin{tabular}{c *{4}{d}}
 & \mbox{C$_6$} \footnote{Units are kJ/mol for energy and nm for distances.} & \mbox{C$_{12}$}
 & \mbox{$\epsilon_{\text{min}}$} & \mbox{R$_\text{min}$} \\
\hline
CH$_4$:OW & 5.873168 $$\cdot10^{-3}$$ & 9.514026 $$\cdot10^{-6}$$ & 0.906 & 0.385 \\
CF$_4$:OW & 1.34270763 $$\cdot10^{-2}$$ & 4.967905 $$\cdot10^{-5}$$ & 0.907 & 0.441 \\
Na$^+$:OW & 4.3426 $$\cdot10^{-4}$$ & 2.3523 $$\cdot10^{-7}$$ & 0.200 & 0.320 \\
Cl$^-$:OW & 6.0106 $$\cdot10^{-3}$$ & 1.6778 $$\cdot10^{-5}$$ & 0.538 & 0.421 \\
SPC/E OW:OW & 2.6173456 $$\cdot10^{-3}$$ & 2.634129 $$\cdot10^{-6}$$ & 0.650 & 0.355 \\
TIP3P OW:OW & 2.4889 $$\cdot10^{-3}$$ & 2.4352 $$\cdot10^{-6}$$ & 0.636 & 0.354
    \end{tabular}
  \end{ruledtabular}
\end{table*}
.

\newpage
\begin{table}
  \caption{Gaussian approximation error width (kJ/mol) from TIP3P water $\mu^{\mathrm{ex}}_{\mathrm{OS,LR}}$ calculation, and bound widening terms.}
  \label{tbl:mferr}
  \begin{ruledtabular}
    \begin{tabular}{ *{3}{r}}
$\lambda$ (\AA) 
& \mbox{2$^\text{nd}$ Order Correction} \footnote{$\tfrac{\beta}{4} ({\left\langle\delta \Delta U^2\right\rangle}_{\lambda+\Delta U}-{\left\langle\delta \Delta U^2\right\rangle}_{\lambda})$}
& \mbox{Bound Widening}\footnote{$\tfrac{\beta}{2} (W_{MM}+V_{MM})$} \\
\hline
2.6 & 15.9 & 0.03 \\
2.7 & 10.7 & 0.06 \\
2.8 & 11.6 & 0.07 \\
2.9 &  7.7 & 0.08 \\
3.0 &  6.8 & 0.08 \\
3.1 &  5.3 & 0.09 \\
3.2 &  4.0 & 0.10 \\
3.3 &  2.7 & 0.09 \\
3.5 &  1.0 & 0.76
    \end{tabular}
  \end{ruledtabular}
\end{table}
.

\newpage
\centerline{\bf FIGURE CAPTIONS}

\noindent
{\bf Figure 1.} Distributions of binding energy of methane to SPC water. The probability distribution of the interaction energy between a one-site methane solute and SPC water (9 {\AA} LJ cutoff). Note that the overlapping region is sparsely sampled (occupying less than 0.2\% of the uncoupled distribution).  The non-interacting data were taken from $2 \cdot 10^7$ particle insertions during a 10 ns solvent-only simulation.  The fully coupled system data were collected from 6000 samples over 20 ns.  Two observed single-count outliers at the right of this distribution are included.

\noindent
{\bf Figure 2.} This figure illustrates the construction of the probability of closest solvent approach, $\func{P}{}{r_{\text{min}}}$. $\func{P}{}{r_{\text{min}}}$ is the probability of observing a cavity of size $r_{\text{min}}$ {\em and} a solvent particle in the shell $r_{\text{min}} \rightarrow r_{\text{min}} + dr_{\text{min}}$.  This is equivalent to the conditional probability for observing a solvent molecule in the prescribed shell, given the existence of the cavity, times the cavity probability.  This physical picture is used to derive the SPT expression for the hard-sphere chemical potential. The inner circle is the unoccupied cavity, while the dashed circle indicates the shell for the observation of solvent closest approach.  The solvent molecule centers are indicated by the small gray circles.    

\noindent
{\bf Figure 3.} Division of the minimum solute--solvent distance ($r_{\mathrm{min}}$) into successive shells.  Derived quantities are shown above the axis (from bottom-up): probabilities of $r_{\mathrm{min}}$ falling in each shell, $s_k$, and cavity probabilities $p_i$.  The logical organization of simulation data is shown below the axis (from top to bottom): bin counts $x_j(k)$ from simulations including hard spheres of size $\lambda_j$, total samples from each simulation, $N_j$, and bin totals $\zeta_k$.

\noindent
{\bf Figure 4.} 
Solute/water oxygen radial distribution functions of all systems studied.  Solid lines (left scale) show 
g(R), while the integrated radial distribution functions, n(R), are shown as dashed lines (right scale).  The RDFs for sodium and chloride can be visually distinguished because of their large size difference.  There the Cl$^-$ dashed line shows g(R) and the dotted line is the corresponding n(R).  The subscripts CO, OO, and IO label the carbon/water-oxygen, water-oxygen/water-oxygen, and ion/water-oxygen distances, respectively.

\noindent
{\bf Figure 5.} 
Solvation free energy contributions for the various systems studied.  In each plot, the solid horizontal line labels the comparison result: test particle insertion for CH$_4$ (9.3 kJ/mol) and CF$_4$ (15.3 kJ/mol), the result of White and Meirovitch\cite{rwhit04} for TIP3P water (-23.43 kJ/mol), and test particle insertion plus charging free energy for the ions (-392.5 kJ/mol for Na$^+$ and -370.0 kJ/mol for Cl$^-$).  The (x) symbol is for the IS contribution, the (+) symbol labels the OS packing component, and the (*) symbol is for the OS long-ranged contribution computed from our mean-field average. The bounds on the OS long-ranged contribution are indicated by dotted lines in each figure, and the BAR result for TIP3P water is given by a line as indicated in the figure.   Triangles label the sum of all free energy contributions as calculated using the present methodology.

\noindent
{\bf Figure 6.} 
Effect of conditioning on the interaction energy distribution.  The upper left figure contains no conditioning. The remaining figures are labeled with the hard-particle conditioning radii.  TIP3P interaction energy distributions become increasingly Gaussian as $\lambda$ increases from zero to 3.5{\AA}  -- evidenced by plotting the distributions corresponding to the coupled (x) and uncoupled (+) cases.

\noindent
{\bf Figure 7.} 
Bayesian estimation of the $\mu^{\mathrm{ex}}_{\mathrm{IS}}$ and $\mu^{\mathrm{ex}}_{\mathrm{OS,HS}}$ profiles for TIP3P water.  Each set of lines with the same style shows the mean plus or minus one standard deviation, leading to upper and lower estimates of the free energy.  In the top panel, $\mu^{\mathrm{ex}}_{\mathrm{OS,HS}}$ was calculated using all 4.5 ns of data from model potential simulations with radii 0 (1$^\text{st}$ branch), 2.7 (2$^\text{nd}$), and 3.3 \AA (3$^\text{rd}$).  Similarly, $\mu^{\mathrm{ex}}_{\mathrm{IS}}$ used radii of 0, 3.0, and 3.2 \AA.  In the lower panel, only the last 200 ps of data from simulations with model potential radii of 0, 2.6, 2.9, and 3.2 {\AA} were used to calculate $\mu^{\mathrm{ex}}_{\mathrm{OS,HS}}$ and radii of 0, 2.9, 3.0, 3.2, and 3.3 {\AA} for $\mu^{\mathrm{ex}}_{\mathrm{IS}}$.

\newpage
\begin{figure}
\includegraphics[angle=-90,width=0.9\linewidth]{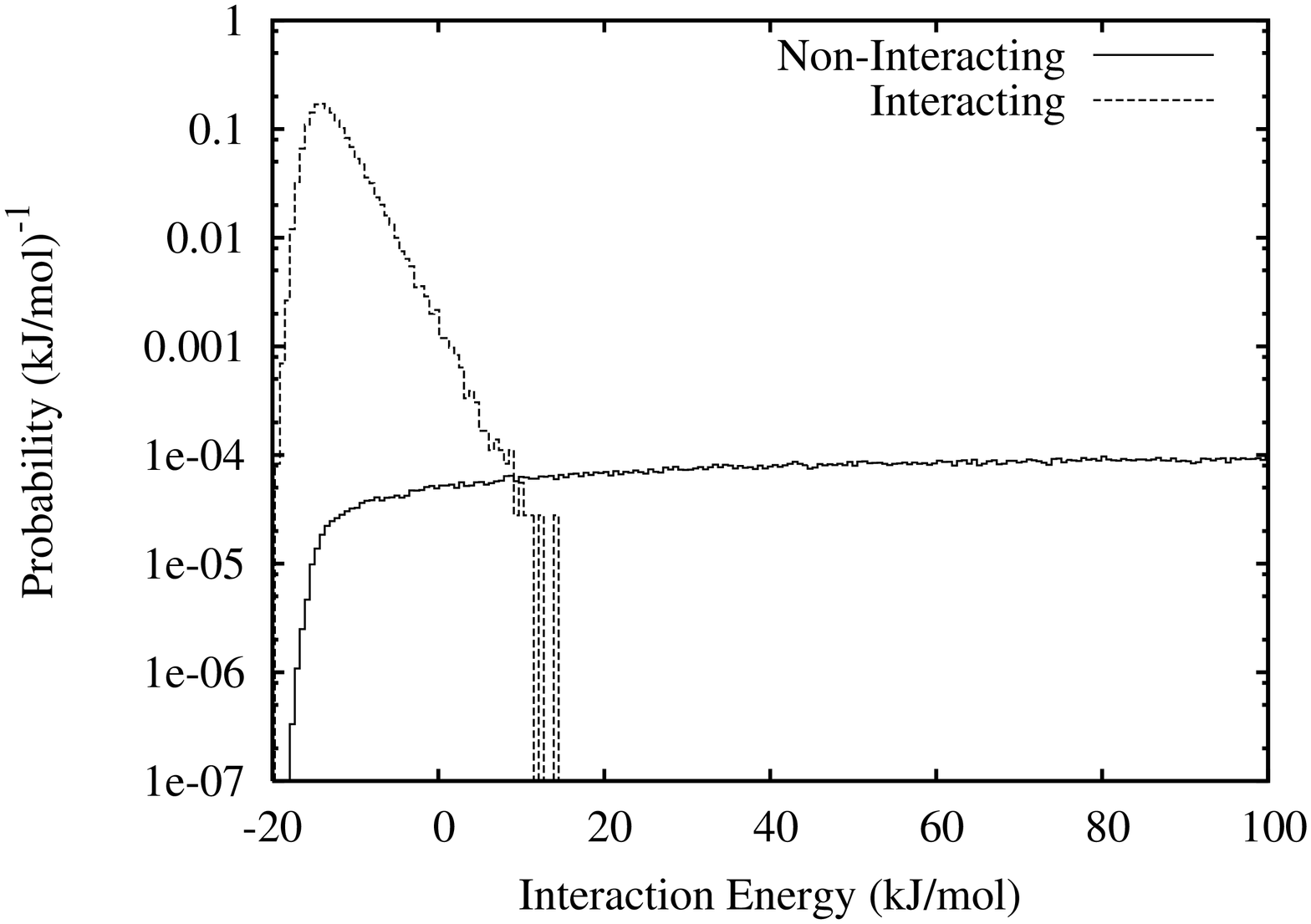}
\caption{}
\label{fig:attpi}
\end{figure}
.

\newpage
\begin{figure}
\includegraphics[angle=0,width=0.7\linewidth]{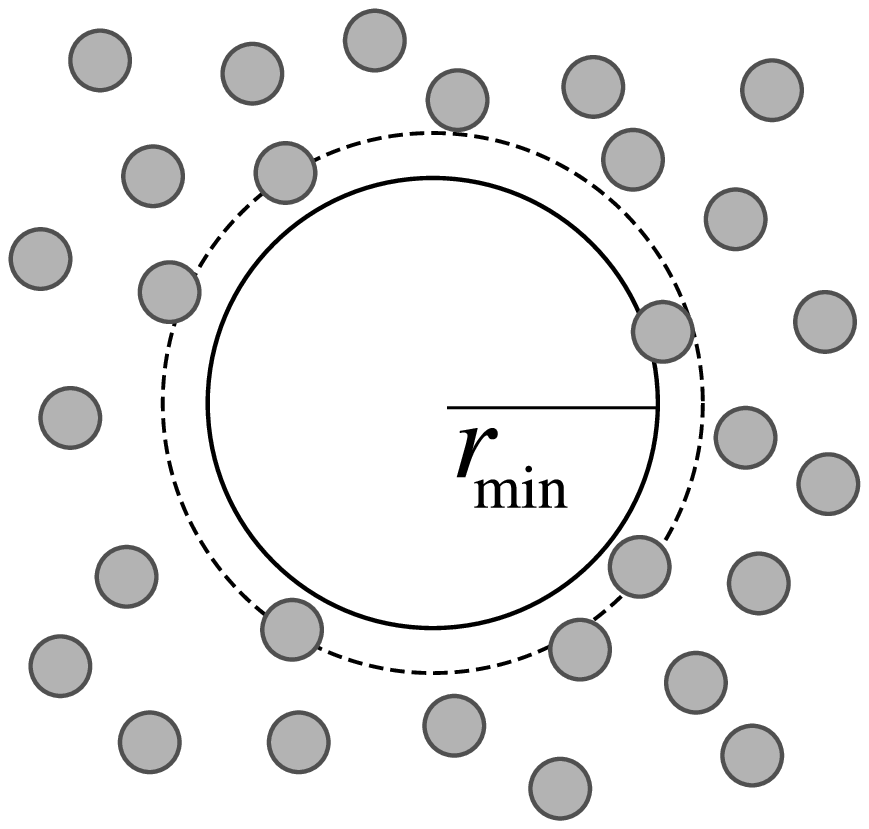}
\caption{}
\label{fig:hertz}
\end{figure}
.

\newpage
\begin{figure}
\includegraphics[angle=0,width=0.9\linewidth]{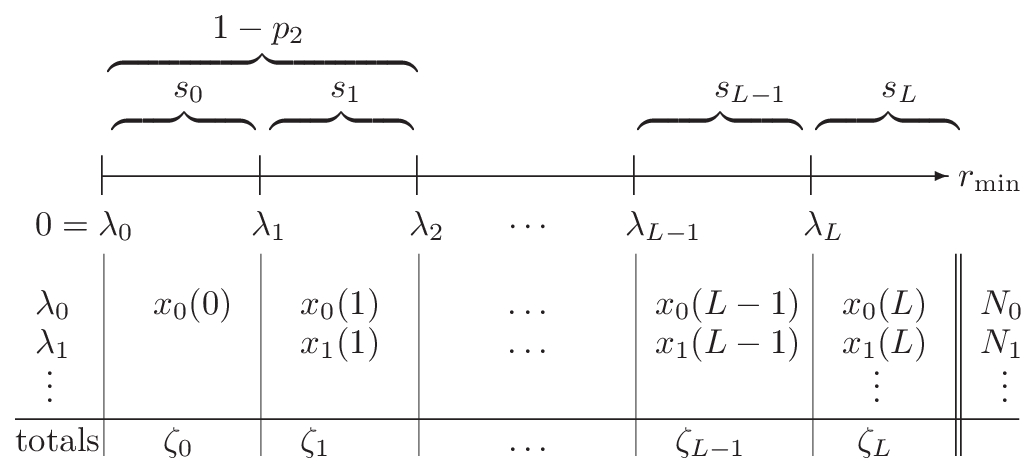}
\caption{}
\label{fig:schem}
\end{figure}
.

\newpage
\begin{figure}
\centering
\includegraphics[angle=0,width=6.5in]{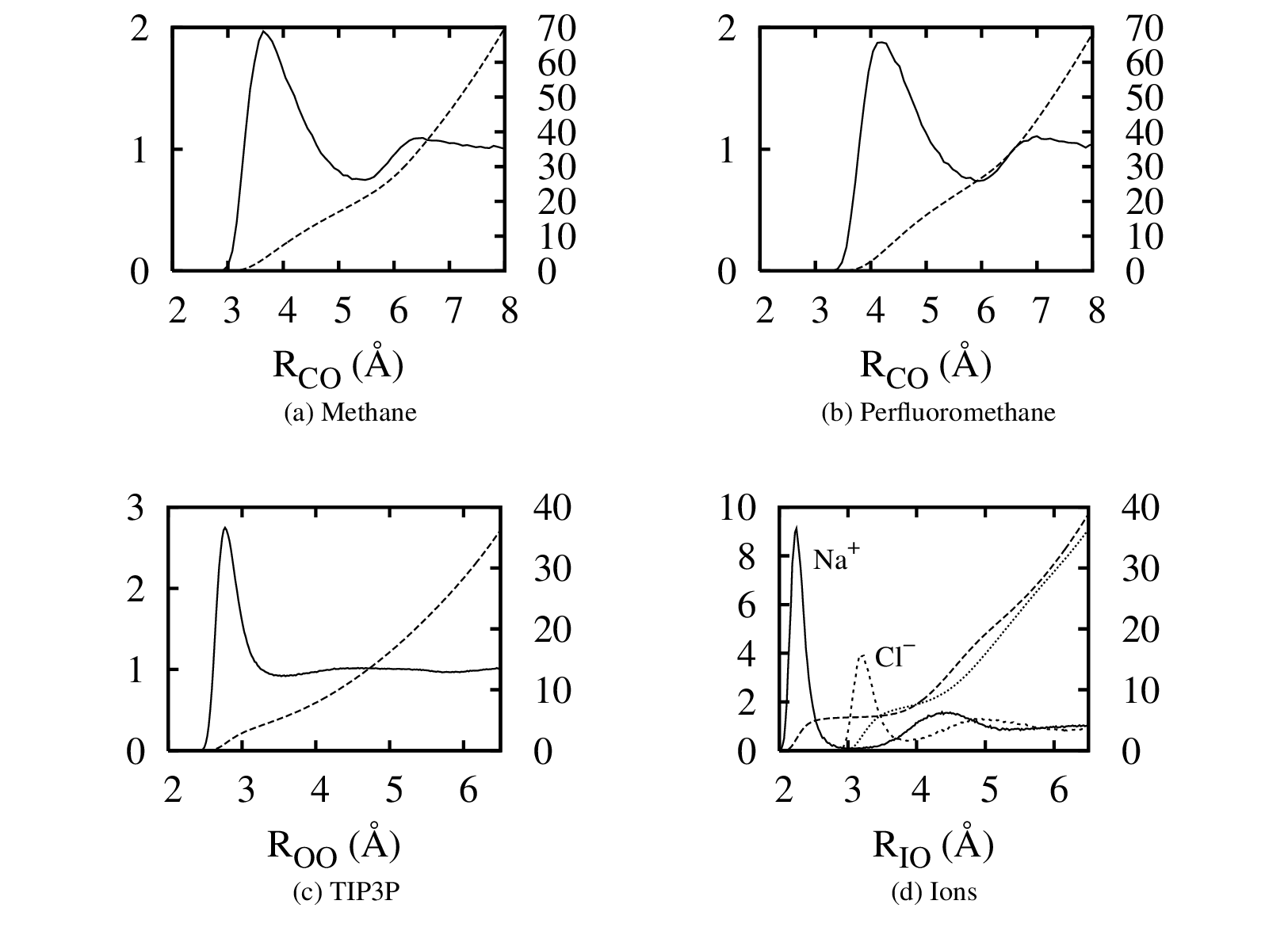}
\caption{}
\label{fig:RDFplots}
\end{figure}
.

\newpage
\begin{figure}
\centering
\includegraphics[width=6.5in,angle=0]{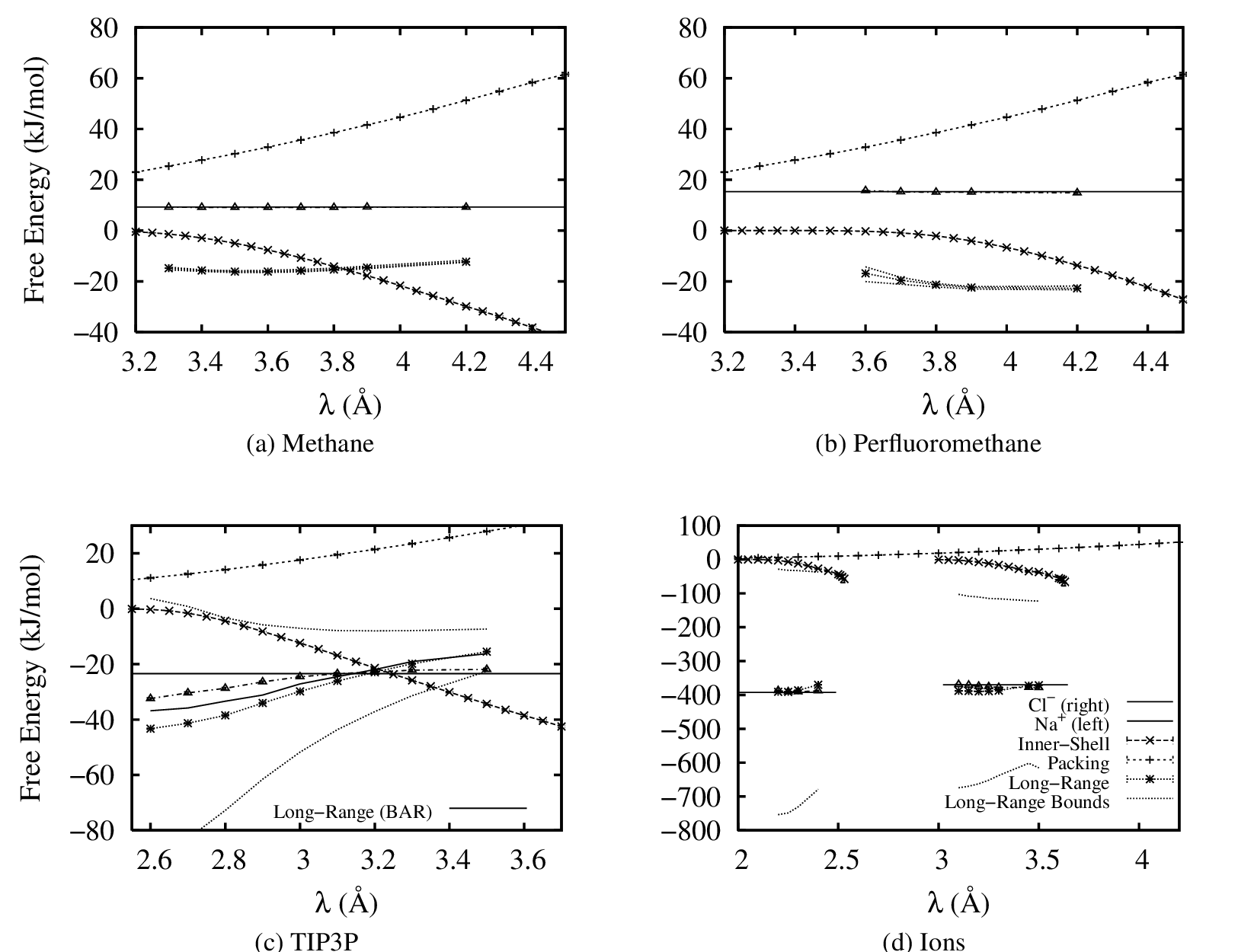}
\caption{}
\label{fig:FEplots}
\end{figure}
.

\newpage
\begin{figure}
\includegraphics[angle=-90,width=0.9\linewidth]{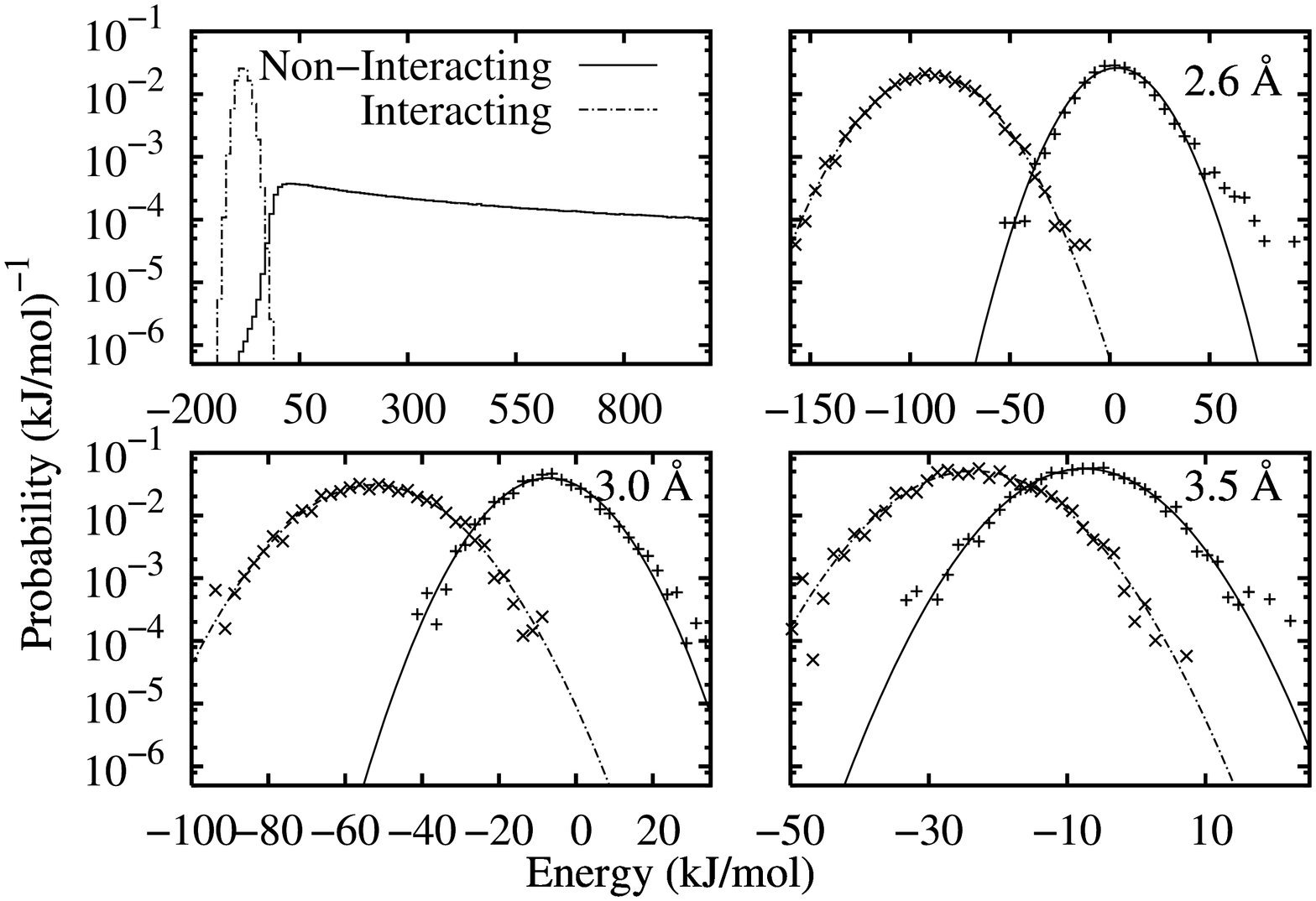}
\caption{}
\label{fig:overlap}
\end{figure}
.

\newpage
\begin{figure}
\includegraphics[angle=-90,width=0.45\linewidth]{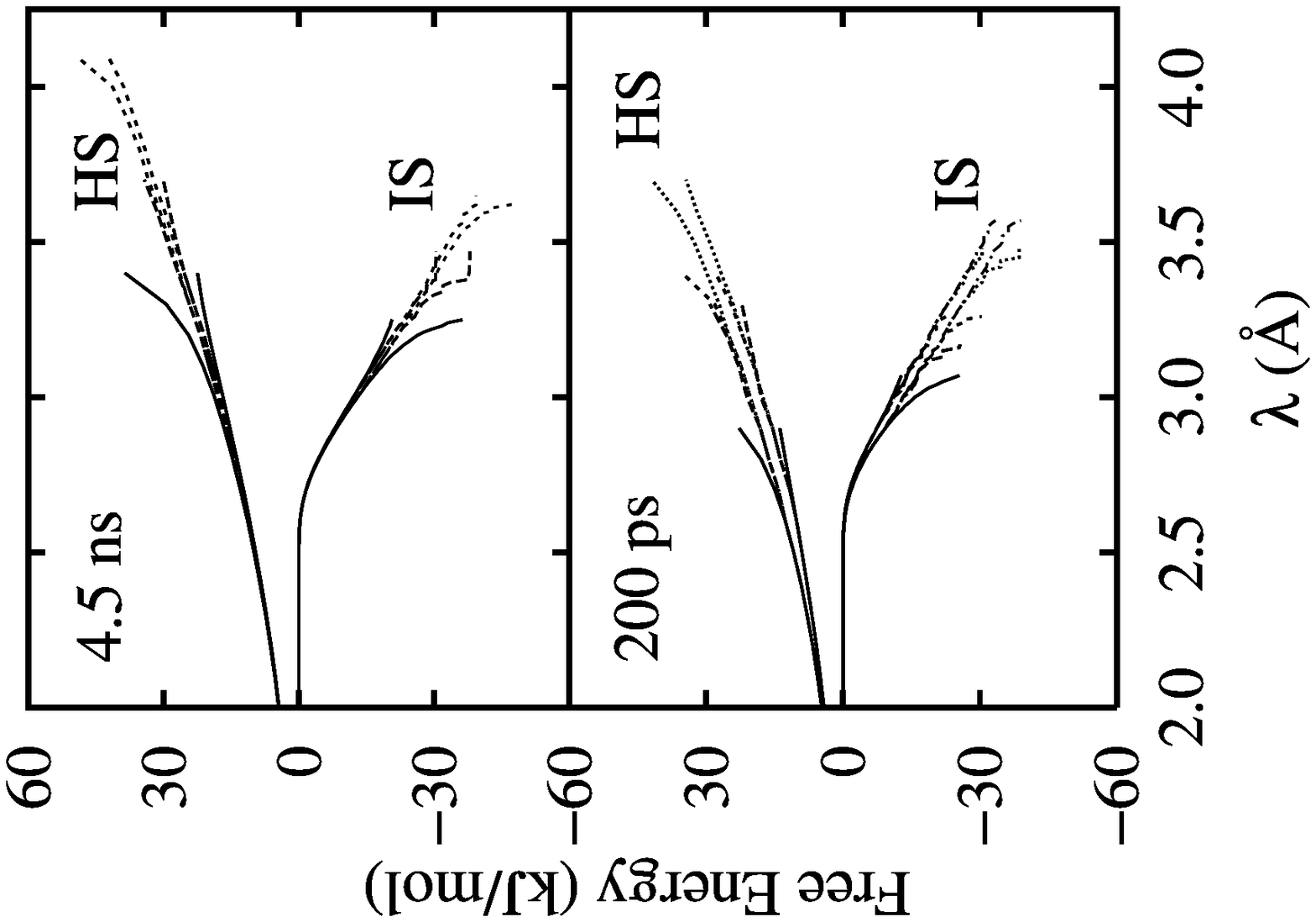}
\caption{}
\label{fig:build}
\end{figure}

\end{document}